# Initial validation of a soil-based mass-balance approach for empirical monitoring of enhanced rock weathering rates

Tom Reershemius[1*] and Mike E. Kelland[2*], Jacob S. Jordan[3], Isabelle R. Davis[1,4], Rocco D'Ascanio[1], Boriana Kalderon-Asael[1], Dan Asael[1], T. Jesper Suhrhoff[5,1], Dimitar Z. Epihov[2], David J. Beerling[2], Christopher T. Reinhard[6], Noah J. Planavsky[1,5]

[1]Department of Earth and Planetary Sciences, Yale University, New Haven, CT, USA
[2]Leverhulme Centre for Climate Change Mitigation, School of Biosciences, University of Sheffield, Sheffield, UK
[3]Porecast Research, Lawrence, KS, USA
[4]School of Ocean and Earth Science, University of Southampton Waterfront Campus, Southampton, UK
[5]Yale Center for Natural Carbon Capture, Yale University, New Haven, CT, USA
[6]School of Earth and Atmospheric Sciences, Georgia Institute of Technology, GA, USA

*corresponding and equal contribution authors: tom.reershemius@yale.edu, mekelland1@sheffield.ac.uk.

**Synopsis**: We describe and empirically validate a geochemical mass-balance approach for tracking rates of Enhanced Rock Weathering in soils, a necessary step in scaling up a promising Carbon Dioxide Removal strategy.

**Abstract**: Enhanced Rock Weathering (ERW) is a promising scalable and cost-effective Carbon Dioxide Removal (CDR) strategy with significant environmental and agronomic co-benefits. A major barrier to large-scale implementation of ERW is a robust Monitoring, Reporting, and Verification (MRV) framework. To successfully quantify the amount of carbon dioxide removed by ERW, MRV must be accurate, precise, and cost-effective. Here, we outline a mass-balance-based method where analysis of the chemical composition of soil samples is used to track *in-situ* silicate rock weathering. We show that signal-to-noise issues of *in-situ* soil analysis can be mitigated by using isotope-dilution mass spectrometry to reduce analytical error. We implement a proof-of-concept experiment demonstrating the method in controlled mesocosms. In our experiment, basalt rock feedstock is added to soil columns containing the cereal crop *Sorghum bicolor* at a rate equivalent to 50 t ha$^{-1}$. Using our approach, we calculate rock weathering corresponding to an average initial CDR value of 1.44 ± 0.27 tCO$_2$eq ha$^{-1}$ from our experiments after 235 days, within error of an independent estimate calculated using conventional elemental budgeting of reaction products. Our method provides a robust time-integrated estimate of initial CDR, to feed into models that track and validate large-scale carbon removal through ERW.

# 1 Introduction

Avoiding 2°C of global warming by 2100 will require dramatic carbon emissions reduction, meaning governments must implement policies with increasingly stringent year-on-year carbon mitigation targets (1, 2). Even the full implementation of all emissions mitigation policies, as of 2022, will result in a 12 gigatonne ($10^9$ tonne) $CO_2$-equivalent shortfall to climate goals as outlined by the Paris Agreement (3). In the absence of feasible pathways to sufficiently reduce carbon emissions, large-scale Carbon Dioxide Removal (CDR) will likely be essential for augmenting decarbonization efforts in the coming century (e.g., 4,5).

Enhanced Rock Weathering (ERW) is a promising CDR technique where naturally occurring mineral weathering reactions that consume atmospheric $CO_2$ are accelerated. This may be achieved by applying crushed silicate rocks with a high reactive surface area to agricultural and forest soils (e.g., 6-21). Potential advantages and co-benefits of ERW include a low technological barrier to implementation at scale (8, 13); long-term storage of carbon compared to organic reservoirs (>10,000 years) (22-28); and supply of key nutrients for crop growth (13, 29-37). Additionally, ERW feedstocks such as basalt may be used for the deacidification of soils, filling the role of agricultural lime (currently a net source of $CO_2$ to the atmosphere; 38-46). Several recent studies have improved our understanding of ERW: mechanistic modelling of weathering reactions in agricultural soils (e.g., 11, 14, 47, 48); modelling hydrological effects on weathering rates (e.g., 49); laboratory and mesocosm experiments tracking uptake of nutrients by plants, and feedstock dissolution rates (e.g., 31, 33, 46, 50-56); and field experiments implementing ERW at scale (e.g., 35, 57-60).

Despite these recent advances, ERW currently lacks a robust and widely accepted framework for Monitoring, Reporting, and Verification (MRV) of CDR rates. This represents a significant barrier to widespread implementation of ERW, in voluntary or compliance markets, or as a subsidized agronomic practice. There will be strong variability in the rates of rock weathering in agricultural settings with variable hydroclimatic conditions, highlighting the need for empirical constraints on weathering at this stage (e.g., 61). Approaches to MRV that are entirely model-based are yet to be fully validated for ERW. The current generation of reactive transport models for simulating ERW have proven useful for making testable predictions (e.g., 14, 19, 47, 48). However, it is not yet clear whether such models are capable of accurately predicting CDR at deployment scale. Therefore, any modelling approach to estimate CDR through ERW at scale must have at its center a robust empirical MRV framework from a diverse set of environments to impart confidence to key stakeholders. The MRV framework must successfully report site- and time-specific rates of feedstock weathering while being cost-effective and minimally invasive (e.g., 62, 63).

In order to quantify weathering rates and/or initial CDR rates from ERW experiments, previous mesocosm and field studies have used measurements of soil inorganic carbon (e.g., 33, 55, 57, 59); the concentration of dissolved ions in porewaters and effluent waters, including cations such as $Ca^{2+}$ and $Mg^{2+}$, as well as carbonate alkalinity (e.g., 31, 33, 50-54, 56, 58, 59, 64); and Sr, Li, Mg, and C isotopic analysis of waters, soils, and rocks (e.g., 33, 51, 52, 59, 64). These data provide a valuable insight into the rate of weathering of feedstocks and the fate of reaction products at different stages of transport from topsoil to the river-ocean system, and it is therefore important that such measurements be made for a representative range of ERW deployment scenarios.

However, basing a site-specific empirical MRV protocol on these metrics is challenging: significant carbonate precipitation in soils is unlikely to occur within the pH range of most agricultural soils that require amendments for deacidification (as well as being undesirable); and collecting porewater and effluent water for analysis of field-specific weathering rates is time- and labor-intensive, and at watershed scale is only feasible in limited settings, such as zero-order streams (e.g., 59). For these reasons, such measurements likely cannot be used in every ERW deployment, if the technology is to scale.

Recent work using measurements of soil exchangeable cations and electrical conductivity as proxies for weathering and alkalinity generation respectively have been welcome steps toward building an MRV toolkit that meets the criteria for providing an empirical base in a wide range of agricultural ERW deployments, by tying in to existing agronomic practices or introducing practices that can be easily scaled (60, 65). However, there can be large errors associated with tracking alkalinity fluxes through electrical conductivity (see ref. 65, 66), while calculating weathering rates purely from the size of the soil exchangeable fraction presents a minimum weathering estimate.

Here, we add to this toolkit by introducing and providing an initial proof of concept study of a soil-based mass balance approach for quantitatively tracking enhanced rock weathering in soils. This approach measures the difference in concentration of ERW feedstock within a soil *in-situ* before and after weathering, directly building from techniques widely used to gauge the extent and mode of weathering in natural systems (e.g., 67-79). We compare the concentrations of mineral-bound metal cations ($Ca^{2+}$, $Mg^{2+}$, $Na^+$) in the solid phase of soils before and after feedstock deployment. We do this by estimating changes in the total amount of these metal cations (CAT) in a specific soil sample relative to the concentration of an immobile tracer, in this case titanium (Ti). Hereafter we refer to this method as TiCAT. As a first step towards robust validation, we compare estimates of the extent of *in-situ* basalt feedstock dissolution from TiCAT to rates determined independently from detailed pool and flux tracking (>2000 measurements) in a mesocosm experiment (following a similar method to that in ref. 33). We then discuss the practical considerations for this approach to be scaled for industrial scale deployment of CDR through ERW, and the steps require to move from an empirical estimate of feedstock dissolution rates to robust error-bounded estimates of CDR.

## 2 Materials and Methods

### 2.1 Theoretical Basis

TiCAT is a mass balance approach for estimating the time-integrated amount of weathering of a rock feedstock – in this case basalt – within a soil sample. This method builds on approaches for estimating the extent of weathering in natural systems, where the concentrations of mobile major cations in an unweathered parent material are compared to those in an equivalent amount of weathered material. The concentration of an immobile trace element is used to establish this equivalence (e.g., 67-79; see also *Supporting Information Section 1.9*). In the TiCAT framework, the unweathered parent material (basalt feedstock) is mixed into the soil. This presents a challenge, as in a field setting the amount of basalt present in a soil sample taken after deployment will not necessarily be proportional to the total amount of basalt deployed, given soil mixing may not be perfectly homogeneous, and some erosion may occur.

To calculate the amount of unweathered parent material initially present in a sample taken after weathering of some of this material has occurred, we first compare soil samples after basalt amendment and weathering with soil samples representative of a pre-amendment baseline, as well as samples of the initial basalt feedstock. This can most readily be visualized as simple two-component mixing between a soil ($c_s$) and a basalt ($c_b$) endmember (**Fig. 1a**). We use the difference in concentration of an immobile trace element, $i$ (e.g., Ti, which is widely used for this purpose, see 70, 80), between a post-application sample ($c_{end}$) and the pre-application soil baseline, to calculate the amount of basalt that has been added to the original soil for the specific sample analyzed ($[Ti]_{end} - [Ti]_s = [Ti]_{add}$). The assumption of immobility is critical to this approach, so while other elements could be used instead of Ti, it must be demonstrated that this criterion is met (see *Supporting Information Section 1.9*).

Using the relative abundance of Ti and mobile major cations (CAT) in the original basalt feedstock, we can then calculate the corresponding amount of mobile major cations, $[CAT]_{add}$, from the basalt feedstock present in the soil + basalt mixture at the point of basalt application (**Fig. 1b**). For a generic cation, CAT:

$$[CAT]_{add} = [Ti]_{add} \times \frac{[CAT]_b - [CAT]_s}{[Ti]_b - [Ti]_s}, \qquad (1)$$

Subtracting the amount of cation in the post-application sample, $[CAT]_{end}$, and adding the soil baseline, $[CAT]_s$, we can calculate the difference in the amount of the mobile cation, $\Delta[CAT]$, between the expected amount from addition of basalt, and the observed amount in the soil + basalt mixture after weathering (**Fig. 1c**):

$$\Delta[CAT] = [CAT]_{add} + [CAT]_s - [CAT]_{end} \quad . \qquad (2)$$

$\Delta[CAT]$ therefore represents the amount of cation loss due to basalt dissolution (i.e., exported from the solid phase) between the point of basalt application and the post-application sampling date. We can also define basalt dissolution as a fraction, $F_D$, where:

$$F_D = \frac{\Delta[CAT]}{[CAT]_{add}} \quad . \qquad (3)$$

The concentration of Ti and CAT in samples may be affected by basalt dissolution reducing the mass of the system. Therefore, a correction for mass loss is applied to $F_D$ (see *Supporting Information Section 1.9*). Multiplying $F_D$ for each cation by the application rate of basalt-hosted CAT gives a cation-specific estimate for weathering of the basalt feedstock at the application scale, assuming that the extent of weathering in an individual sample is representative. Variability in hydrology and soil characteristics (e.g., 81) means that at field-scale, it is likely that multiple samples from sites representative of a range of field conditions (pH, density, etc.) will need to be analyzed for a representative weathering estimate to be calculated.

From the calculated *in-situ* cation-specific weathering rates of basalt feedstock, an initial rate of CDR (i.e. conversion of carbonic acid to bicarbonate) can be calculated, assuming the acidity consumed during silicate mineral dissolution is ultimately sourced from atmospheric $CO_2$ (see

*Supporting Information Section 1.9* for more details). In many agricultural settings fertilizer amendments such as urea ammonium nitrate (UAN) are used. Nitrification of reduced nitrogen species is a source of strong acid that can also contribute to mineral weathering. In this case, feedstock weathering may or may not result in CDR. In cases where the strong acid would have interacted with a silicate mineral already, this weathering needs to be discounted from initial CDR estimates (see *Supporting Information Section 1.10*). The initial CDR rates calculated can be thought of as maximum possible CDR from an ERW deployment, and do not take into account downstream processes that will reduce the efficacy of carbon storage as bicarbonate in the river-ocean system, such as rerelease of $CO_2$ via the precipitation of secondary clays and carbonates outside the soil column, or potential degassing of $CO_2$ after conversion of bicarbonate to carbonic acid, from reequilibration in acidic solutions (e.g., 20, 63, 82-84).

## 2.2 Analytical requirements

Resolvability of a dissolution signal from soil-based mass-balance is a function of analytical uncertainty, feedstock application rate, and extent of feedstock dissolution. Limiting uncertainty on instrumentation used to analyze elemental composition is a critical aspect of the TiCAT approach — and is likely to be a critical issue for any other approach aiming to track CDR from the solid phase. This is due to the signal-to-noise ratio associated with measuring a small amount of feedstock mixed into a large amount of background soil. For technically and commercially feasible feedstock application rates (likely <50 t ha$^{-1}$, see 13), analytical uncertainty can result in overlap between error-bounded values for cation concentration of soil-feedstock mixtures before and after dissolution has occurred. Using representative soil and basalt compositions from this study for instance, at 5% analytical uncertainty (a typical lower bound for global analytical uncertainty in X-ray fluorescence measurements of major element concentration in soils, see e.g., 85-87, and *Supporting Information Section 1.7*) a 25% loss of major cations from the basalt portion of a soil-feedstock mixture is unresolvable even at an application rate of 100 t ha$^{-1}$, assuming the mixture is homogenized to a depth of 10cm (a common mixing regime for managed row crop systems). However, at 1% analytical uncertainty the same extent of cation loss is resolvable at ~25 t ha$^{-1}$ (**Fig. 2**). Thus, when accounting for a range of plausible dissolution and application rates, analytical error must generally be limited to ~1% for our mass balance method to be accurate and widely applicable, a standard that is more stringent than that currently achievable by most commercial inorganic elemental analysis, including commercial mass spectrometry (e.g., 88, 89).

We obtain the requisite analytical precision for applying the TiCAT method here using isotope dilution inductively coupled plasma mass spectrometry (ID-ICP-MS). Isotope dilution is a well-established analytical method where the concentration of an element in a sample can be measured from the known concentration of an element in a spike solution, and the ratios of two isotopes of the same element in the natural sample and the spike respectively (90-92). The amount of an element in the sample, $n_{sam}$, is given by:

$$[n]_{sam} = [n]_{spk} \frac{R_{spk} - R_{mix}\Sigma_i R_{isam}}{R_{mix} - R_{sam}\Sigma_i R_{ispk}} \quad . \qquad (4)$$

where R is the ratio of two isotopes, and $\Sigma_i R_i$ is the sum of ratios of all isotopes to a reference isotope (e.g., 90; see also 91, 92). We used an isotope spike 'cocktail', doped with isotopes of Mg,

Ti, and Ca found in lower abundance in natural samples (**Supplemental Fig. S5**). Isotope spikes were prepared from powders of spiked $TiO_2$, MgO and $CaCO_3$. The pure spike Ca carbonate powders were digested using HCl, and the Mg and Ti oxide using $HNO_3$+HCl+HF. Following the digest each spike solution was calibrated by measuring the relative concentration of Mg, Ca, and Ti isotopes on a Thermo Scientific Neptune Plus multicollector ICP-MS for ~48 hours. Estimates of the uncertainty on the spike determination were < 0.1 ‰ based on replicate analysis. Individual spike solutions were then used to make an isotope spike 'cocktail' solution containing Mg, Ca and Ti spikes. The 'cocktail' was added to each sample during the dissolution stage of sample preparation to ensure sample-spike equilibration.

Isotope dilution allows for sample-specific calculation of element concentrations, unlike a standard calibration curve method whereby standard solutions of known concentration are run to relate element intensities (in counts per second) to concentration. Isotope dilution therefore corrects for matrix effects, mass bias, and instrument drift, and thus improves both accuracy and precision of measured concentrations, significantly reducing global analytical uncertainty (**Fig. 3**; see also 93). In addition to an iCAP TQ ICP-MS used to run samples for this study, we analyzed certified reference materials (BHVO-2 basalt and SGR-1b shale) on other ICP-MS models in order to test the data quality achieved by a variety of widely available instruments (a Perkin Elmer NexION 5000 Multi-Quadropole ICP-MS and a Thermo Scientific ElementXR High Resolution Magnetic Sector ICP-MS). For sample runs on the iCAP TQ ICP-MS, we were able to achieve average analytical uncertainty on reference materials of 0.22% for Ti, 0.78% for Mg, 0.39% for Ca, and 0.58% for Na (not using isotope dilution), when calculating uncertainty as the mean difference of calculated to certified values (global analytical uncertainty); and 0.75% for Ti, 1.16% for Mg, 1.29% for Ca, and 2.54% for Na (not using isotope dilution), when calculating uncertainty by the standard deviation of measurements as a percentage of the mean. Our results show that isotope dilution allows for a level of measurement accuracy and precision on quadrupole ICP-MS instruments that would otherwise typically only be achievable with a magnetic sector instrument. Given its much lower cost and far greater availability, the ability to leverage quadrupole ICP-MS for rapid, high-throughput analyses may ultimately be a critical factor in making this MRV technique economically viable at scale.

## 2.3 Mesocosm ERW experiments

As an initial test of the TiCAT method we employed laboratory mesocosm ERW experiments, which allowed us to independently estimate ERW and CDR by measuring the concentration of reaction products in plant, soil exchangeable fraction and leachate solution pools (see 29, 31, 33, 50-54, 56, 59). Each mesocosm contained a single C4 cereal crop *Sorghum bicolor* plant (see 33) with two different fertilizer treatments: nitrogen-phosphorus-potassium (NPK) (n=14), or manure (n=14). To half of the columns for each fertilizer treatment a basalt feedstock was added equivalent to an application rate of 5 kg $m^{-2}$ (50 t $ha^{-1}$), mixed to a depth of 12cm, and left in a controlled environment for 235 days (for a detailed description of mesocosm design and construction, substrate and feedstock preparation and characterization, plant varieties and growth conditions, and irrigation regime, see *Supporting Information*).

Leachate was collected from mesocosms at six discrete leachate events, accounting for the entire leachate flux during the experiment. After 235 days, samples were taken from relevant chemical pools — the soil exchangeable fraction, the solid phase with the exchangeable fraction removed by leaching with ammonium acetate, and the plant material (comprising shoots, roots and seeds). Analysis of total inorganic carbon (TIC) did not show detectable increases in a subset of mesocosm soils tested before leaching with ammonium acetate (using an Eltra C/S Analyzer with detection limit of 0.1wt%), suggesting that carbonate formation should have a negligible impact on the overall cation budget of the mesocosm systems. Aliquots of solid phase samples were then ashed, digested using $HNO_3$+$HCl$+$HF$, and elemental concentrations measured using ID-ICP-MS (see *Supporting Information Section 1.7*). Using a conventional approach focused on reaction products, we calculated elemental budgets for major cations for each mesocosm from the dissolved or non-mineral-bound pools (leachate + soil exchangeable fraction + plant material) (see *Supporting Information Section 1.8*). We compared the elemental budgets calculated for basalt-amended mesocosms to a control mesocosm for each fertilizer treatment that had no basalt applied. Control mesocosms were selected as the control replicates for which the topsoil major element composition most closely matched the initial soil baseline. The excess $Na^+$, $Ca^{2+}$, and $Mg^{2+}$ in the elemental budget of basalt-amended mesocosms relative to the controls were assumed to represent reaction products of basalt dissolution. From the amounts of these cations we obtained initial ERW rates. Using a modified Steinour formulation – a simple stoichiometric approach that relates the amount of mobile cations to the amount of carbonic acidity converted to bicarbonate by charge balance (see *Supporting Information Section 1.3*) – ERW rates were converted to initial CDR estimates.

These rates can be compared with those obtained using the TiCAT method, as the cations released into non-mineral-bound pools from feedstock weathering should correspond to the amount of cation loss from the basalt fraction being weathered. We analyzed the concentration of Na, Ca, Mg and Ti in the solid phase samples taken from the upper 12cm portion of each mesocosm. We calculated $F_D$ of the basalt fraction present in the soil + basalt mixture separately for each major cation, and corrected these for the concentration effect from basalt dissolution (see Section 2.1, also *Supporting Information Section 1.9*). From the corrected cation-specific $F_D$ for each mesocosm and using the application rate of basalt for each mesocosm, we calculated the total amount of major cations in basalt applied over a given area that was dissolved (see *Supporting Information Section 1.9*), giving an initial ERW rate, and using the same modified Steinour formulation as above, an initial CDR estimate. To directly compare the TiCAT method to a weathering product approach, we applied a correction to the TiCAT estimates to account for strong acid weathering from nitrification of fertilizers (see *Supporting Information Section 1.10*).

## 3 Results and Discussion

Our results show agreement between two independent methods of calculating weathering and initial CDR in our mesocosm systems (**Fig. 4a**). The more conventional approach, measuring cation reaction products in dissolved cation pools, yielded mean initial CDR estimates of $1.68 \pm 0.11$ $tCO_2eq$ $ha^{-1}$ (NPK-fertilized) and $1.05 \pm 0.15$ $tCO_2eq$ $ha^{-1}$ (manure-fertilized) across all mesocosms. The TiCAT approach introduced here, which measures cation loss from the solid phase of soil samples, gave mean initial CDR estimates of $1.84 \pm 0.37$ $tCO_2eq$ $ha^{-1}$ (NPK-fertilized) and $1.04 \pm 0.37$ $tCO_2eq$ $ha^{-1}$ (manure-fertilized). Thus, mean initial CDR estimates from the TiCAT method were within error ($\pm$ standard error of means) of those from the reaction product method for both NPK- and manure-fertilized basalt-amended mesocosms.

Mean initial CDR values across all basalt-amended mesocosms were $1.44 \pm 0.27$ tCO$_2$eq ha$^{-1}$ (TiCAT), and $1.36 \pm 0.12$ tCO$_2$eq ha$^{-1}$ (dissolved pools) (**Fig. 4b**). This is broadly consistent with estimated CDR values calculated for similar ERW studies, albeit these range greatly in application amount and duration (see 34). Given a CDR potential for the basalt used in our study of 183.56 kgCO$_2$ t$^{-1}$ (see *Supporting Information Section 1.3*), the initial CDR after 235 days of carbonic-acid-driven weathering was $15.7 \pm 3.1$ % of this potential, using results from TiCAT. Our results thus demonstrate that the solid-phase approach underlying TiCAT produces estimates for initial CDR within error of those calculated by analyzing the dissolved, plant, and soil exchangeable cation pools that constitute the ultimate reaction products in our mesocosm experiments, suggesting that it can yield an accurate and robust estimate of initial CDR in enhanced weathering systems.

It is important to emphasize that the CDR rate estimated based on the time-integrated amount of feedstock dissolution and cation loss should only be regarded as an initial CDR value. There is potential for leakage of initially captured carbon downstream of a given field deployment, as alkalinity and dissolved inorganic carbon are transported from the soil column to the oceans (e.g., 20, 63, 82-84). In addition, a large fraction of the dissolved cation load in any soil will be transiently hosted in soil exchange sites (e.g., 94-97; see also 56). This cation storage at exchange sites is temporary, and upon their release dissolved cations will drive CDR through charge balance in the carbonic acid system (see 51, 52). However, this means there is a variable lag time between feedstock dissolution and CO$_2$ capture that needs to be considered for accurate CDR quantification. Given these factors, a robust, "cradle-to-grave" MRV approach with TiCAT at its core will also require modeling the transport of weathering products through the soil (14, 19, 47, 98) and groundwater-river-ocean system (20, 82-84) to determine potential leakage through re-release of CO$_2$ back to the atmosphere. In the near term, developing and testing these models should be done in conjunction with monitoring of aqueous geochemistry, alongside soil-based approaches (such as TiCAT). As with all CDR techniques, emissions accounting must also be implemented to calculate net CDR rates.

There are several key challenges that need to be met before TiCAT can be widely applied. First of all, it must be demonstrated that scaling from weathering rates at specific sampling points to a larger system allows for a representative measurement of weathering across that system, while minimizing uncertainty. This study suggests that this condition can be met, at least when averaging across mesocosm experiments. Secondly, spatial heterogeneity of elemental concentrations in managed soils must be examined to assess the density and volume of sampling that must be implemented to be able to directly compare between samples of background soil before ERW feedstock amendment and post-amendment soil; in both cases, specific sampling protocols such as pooling and/or gridded sampling may be useful tools in making representative measurements (81), and control sites will be useful for testing these. A complication is that in open systems such as agricultural fields, material may be introduced from external sources that could interfere with the simple two-member mixing model (e.g. flooding events, windblown dust). Further, in some settings other soil amendments may be used in conjunction with silicate minerals. If these contain significant amounts of major cations, or the immobile tracer, a more elaborate mixing model must be used to account for these. Additionally, the TiCAT approach may not be viable as a standalone MRV framework in specific cases: for example in settings with very high physical erosion; in settings where there is significant and fast weathering from soils; where feedstocks are used that have a very similar chemical composition to the soil or have lower concentrations of all immobile trace elements; or where feedstocks are especially slow-weathering, such as in arid settings not conducive to ERW. Other site-specific conditions may require alterations to workflows; for example, the depth of sampling required may vary in response to variable mixing depths, as modulated by tilling practices, crop type, and biological mixing.

Lastly, effective implementation of the TiCAT approach relies on stringent constraints on analytical error, which may be challenging via standard practice with most commercially available measurements for elemental concentration, including ICP-MS (see 88, 89). Nonetheless, here we have demonstrated that it is possible using isotope dilution on a standard quadrupole ICP-MS instrument to minimize analytical error to ~1%, making even <10% basalt dissolution analytically resolvable for total feedstock application rates of 50 t ha$^{-1}$. However, replication of this analytical precision in commercial laboratories will require adjustment to workflows and standard operating procedures. It is also likely that the aggregate impact on unit cost (dollar cost per ton of $CO_2$ captured) of TiCAT as an MRV procedure will broadly follow a "learning curve" trajectory, driving lower costs as ERW is scaled up (e.g., 99).

The TiCAT method overcomes some of the issues with prior methods of estimating ERW, particularly those that rely on accurately measuring the amount and transport of weathering reaction products (i.e., bicarbonate ions, $HCO_3^-$, or cations in soil drainage waters) after feedstock application. Methods that rely on tracking the dissolved phase are extremely time- and labor-intensive, introducing significant barriers to scale. For example, a thorough study of a field-scale ERW trial monitoring aqueous reaction products, such as that conducted by ref. 59, involves many months of labor- and time-intensive sampling not only of soils, plants, and possibly porewaters from the agricultural plots onto which ERW feedstock is applied, but also a wider detailed monitoring of the drainage regime and watershed around such a site. Even using this style of sampling protocol, the necessary granularity of measurements would likely miss short-term fluctuations such as wash-out after rain events, which in many river systems account for an important component of the overall solute discharge (e.g., 100). Such measurements, as well as those reliant on directly measuring soil exchangeable cations, can also be complicated by varying timeframes over which cations are bound to exchangeable sorption sites within a soil (see 51, 52).

Our approach also directly overcomes possibly the largest uncertainty in scaling ERW in agricultural settings — estimating the initial extent of feedstock dissolution in soils (see e.g., 34; 61). There is currently significant uncertainty on how rock grain surface areas evolve through time within a given field setting (i.e., individual farm), and the extent to which secondary mineral formation on the surface of feedstock has the potential to alter mineral dissolution rates (e.g., 101-109). In addition, bulk mineral dissolution kinetics are in some cases poorly constrained (e.g., 105, 110, 111). Taken together, these considerations make it extremely challenging to accurately forecast feedstock dissolution across a range of deployment regimes with existing reactive transport models alone (34, 61, 101-111).

A significant additional advantage to this MRV approach is that it can directly integrate into existing agronomic practices. Samples from the uppermost portion of the soil are already regularly taken for nutrient and soil pH analysis (e.g., 94, 112). Importantly, this means that there is already extensive personnel and infrastructure in place that can be leveraged to scale empirical validation of ERW at minimal cost, in marked contrast to empirical verification of soil organic carbon concentrations (SOC), which requires modified sampling protocols for accurate empirical results. Existing frameworks for carbon storage in agricultural settings are mostly focused on SOC, which does not allow for landowners and land users to include alkalinity generation through practices such as ERW into an estimate of carbon storage. Deployment of ERW at scale requires MRV tools such as TiCAT to be incorporated into these frameworks. This would allow for combined use of ERW and SOC maintenance to achieve maximum carbon storage depending on local conditions.

In summary, we have demonstrated with mesocosm ERW experiments that a soil-based mass balance approach — TiCAT — accurately tracks ERW with a basalt feedstock to allow estimation of CDR rates. TiCAT yields estimates that are within error of those calculated by complete elemental budgeting of weathering reaction products gained in plant and exchangeable cation pools. Using an isotope dilution method, we can reduce analytical error sufficiently that a dissolution signal is resolvable in the solid soil phase at reasonable feedstock application rates. Applying the methods used in this study to field-scale trials is a necessary next step in verifying the capacity of TiCAT to be used for MRV in ERW in the field. Additionally, our approach will ultimately need to be augmented by the development of cradle-to-grave MRV approaches that can provide error-bounded estimates of final CDR. Nevertheless, our results suggest that a soil-based mass-balance method could be a cost-effective and accurate centerpiece of a robust MRV toolkit for deploying ERW at scale.

**Figures**

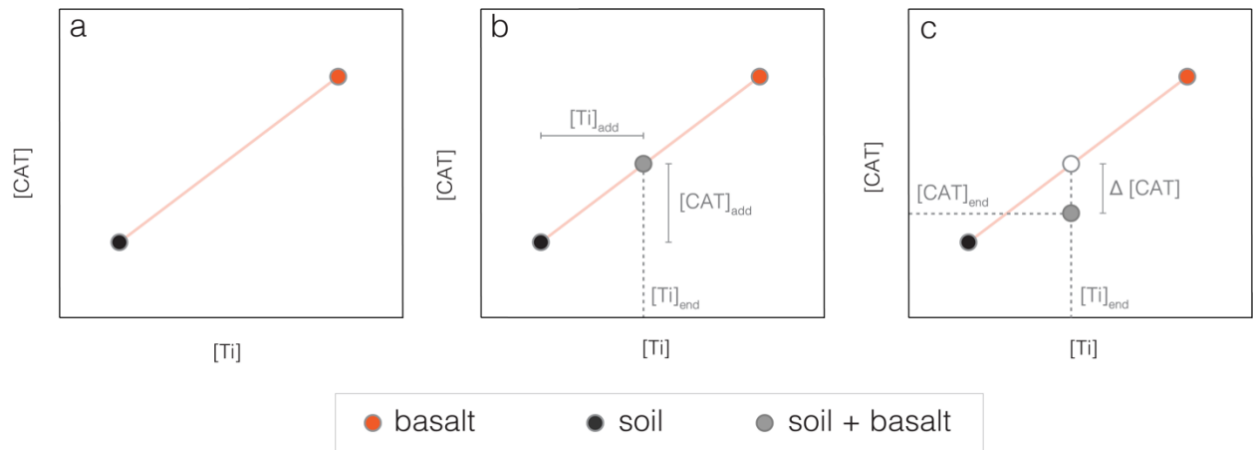

**Fig. 1.** The TiCAT conceptual framework as a simple two-component mixing model. Idealized soil and basalt (a commonly proposed ERW feedstock) endmember compositions are plotted in [Ti] v [CAT] space (a). A mixture of soil + basalt initially plots on the idealized mixing line between both endmembers (b). Dissolution results in loss of [CAT] from the solid phase, while [Ti] is conserved as it is immobile; the original composition of the soil + basalt mixture (indicated by the white circle) is the intersection of $[Ti]_{end}$ with the mixing line, and $\Delta[CAT]$ is the amount of CAT lost by dissolution (c).

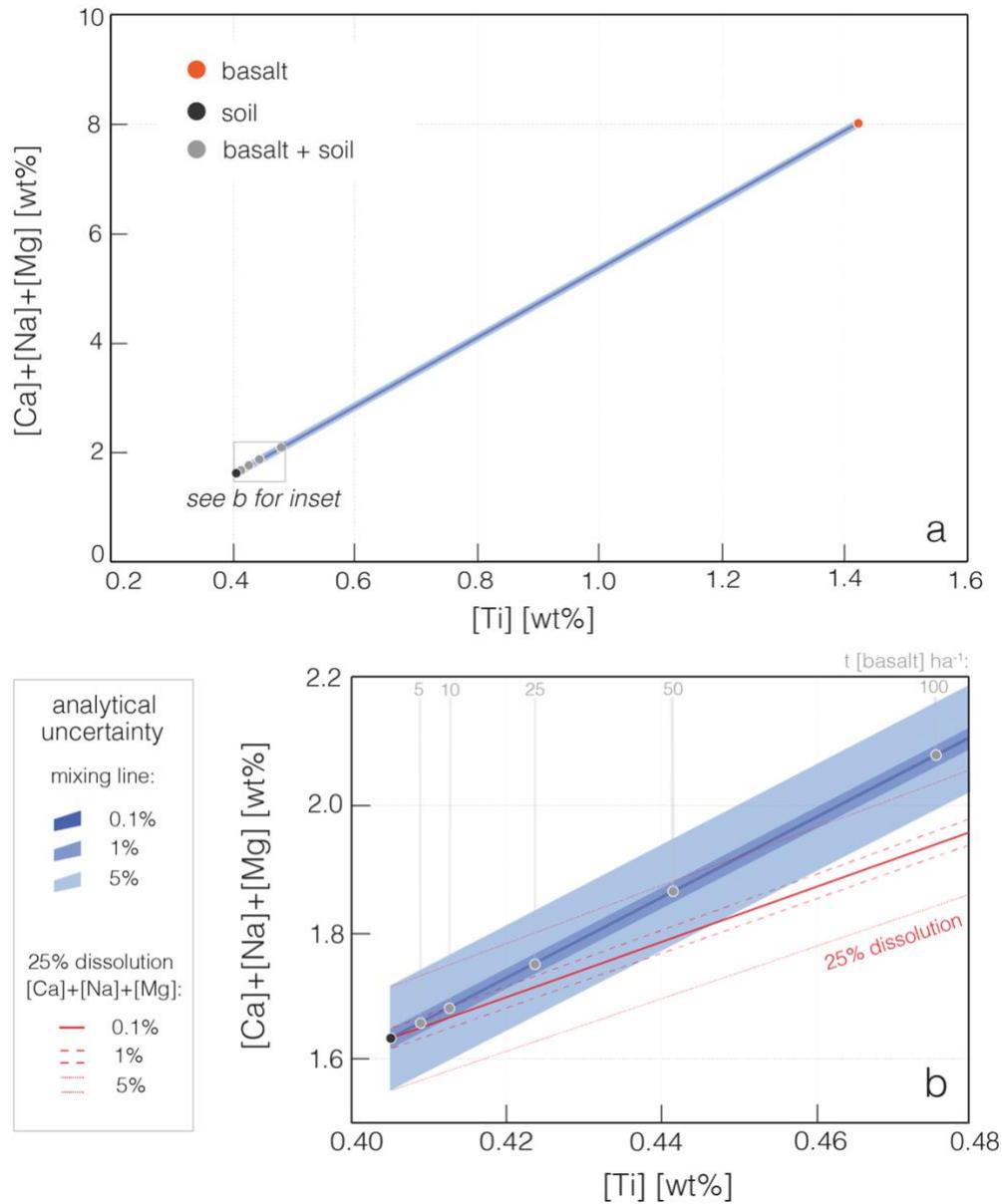

**Fig. 2.** A mixing model with representative data for soil and basalt feedstock endmembers. Assumed homogeneous mixtures of soil + basalt to 10cm mixing depth are shown for a range of basalt application scenarios (see inset, b). Error envelopes are shown for the mixing line and a line indicating theoretical 25% dissolution, based on uncertainty in measuring the elemental concentration of soil and soil + basalt samples. Resolvability of a dissolution signal is dependent on dissolution rate, basalt application rate, and analytical error. Mixing line error envelope assumes that absolute analytical error of basalt is same as soil, a realistic scenario given repeat measurements of a bulk feedstock; dissolution error envelope assumes that sampling gives a representative soil background.

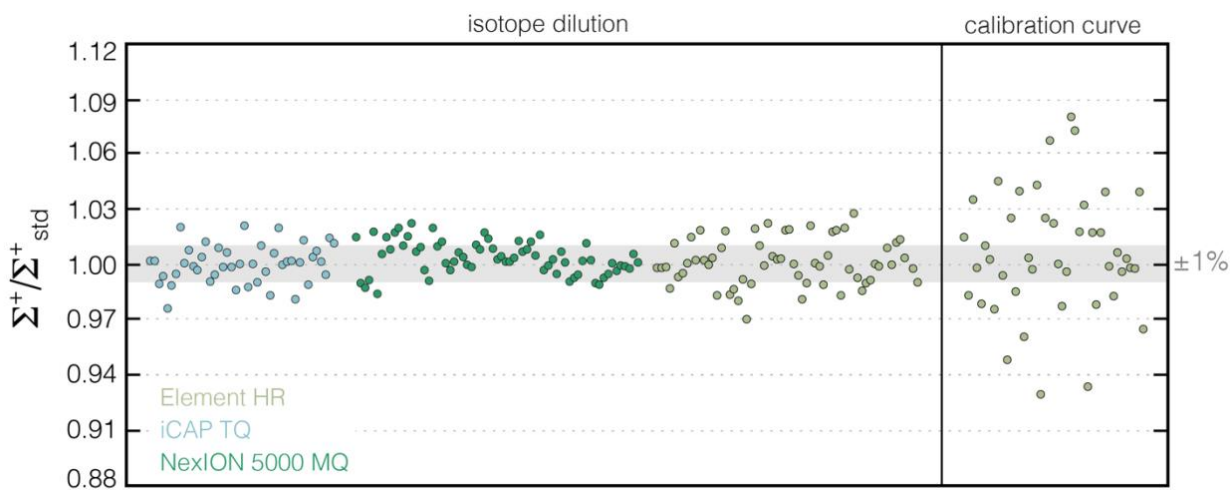

**Fig 3.** Representative analytical error for Ca, Mg, and Ti on standards run on three different ICP-MS instruments, using a calibration curve method and an isotope dilution method. Isotope dilution cannot be applied to Na as it has a single stable isotope. The grey shaded region represents an analytical error range of ±1%. ICP-MS instruments used were two quadropole instruments, a Thermo Scientific iCAP TQ ICP-MS and a Perkin Elmer NexION 5000 Multi-Quadropole ICP-MS; as well as a Thermo Scientific Element High Resolution Magnetic Sector ICP-MS.

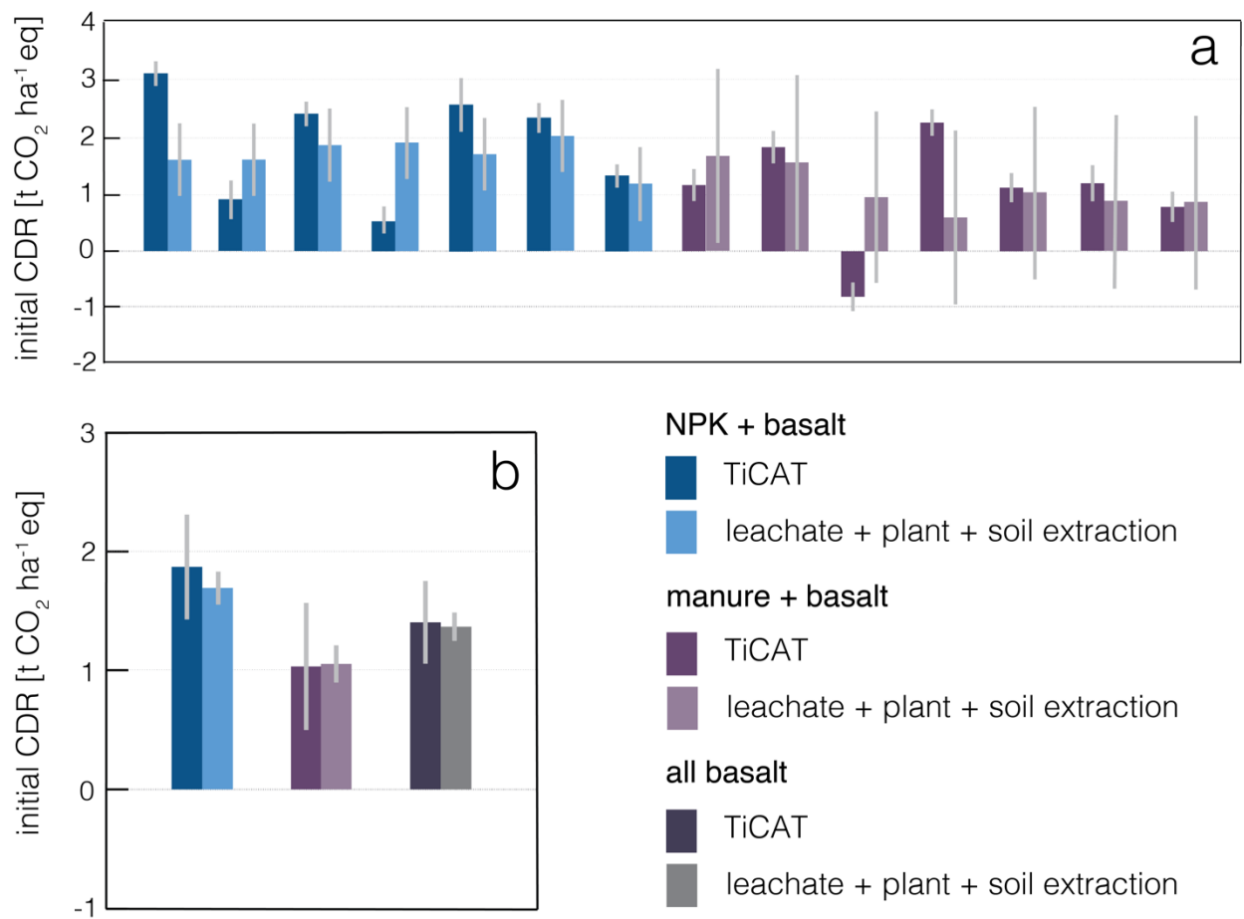

**Fig 4.** Initial CDR estimates calculated using cation data from TiCAT and from non-mineral-bound cation budgets. Results are shown for individual mesocosms (a), and as mean values for all mesocosms pooled by treatment type, where dark grey bars show all basalt-treated mesocosms as a single pooled data set (b). Error bars for TiCAT are propagated analytical error in (a), and standard error of means (± σ) where baseline soil and basalt samples have been pooled. Error bars for leachate + plant + soil extraction are propagated standard error between measurements for all mesocosms of the same treatment in (a), and standard error of means (± σ) where baseline soil and basalt samples have been pooled.

**Supporting Information**

Contains Supporting Materials and Methods (1.1-1.10), Supporting Tables (Tables S1-S2), Supporting Figures (Figures S1-S10), and Supporting References. Supporting Datasets and Software attached (Datasets S1-S3, Software S1).

**Acknowledgments**

We thank Dr Amy L. McBride for providing the basalt mineralogy and particle size distribution data for this study. We also acknowledge technical support from Irene Johnson and Dr Karen J. Bailey during soil excavation, plant tissue preparation and the ammonium acetate soil extraction. Six reviewers gave helpful and constructive feedback that improved this study. N.J.P. and T.J.S. acknowledge support from the Yale Center for Natural Carbon Capture. M.K. acknowledges funding through an NERC/ACCE DTP studentship and D.J.B. and D.Z.E acknowledge funding from the Leverhulme Trust through a Leverhulme Research Centre award (RC-2015-029).

## References

1. Ou, Y.; Iyer, G.; Clarke, L.; Edmonds, J.; Fawcett, A.A.; Hultman, N.; McFarland, J.R.; Binsted, M.; Cui, R.; Fyson, C.; Geiges, A.; Gonzales-Zuñiga, S.; Gidden, M.J.; Höhne, N.; Jeffery, L.; Kuramochi, T.; Lewis, J.; Meinshausen, M.; Nicholls, Z.; Patel, P.; Ragnauth, S.; Rogelj, J.; Waldhoff, S.; Yu, S.; McJeon, H. Can updated climate pledges limit warming well below 2°C? *Science* **2021**, *374*, 693–695. https://doi.org/10.1126/science.abl8976
2. Meinshausen, M.; Lewis, J.; McGlade, C.; Gütschow, J.; Nicholls, Z.; Burdon, R.; Cozzi, L.; Hackmann, B. Realization of Paris Agreement pledges may limit warming just below 2 °C. *Nature* **2022**, *604*, 304–309. https://doi.org/10.1038/s41586-022-04553-z
3. United Nations Environment Programme. *Emissions Gap Report 2022: The Closing Window — Climate crisis calls for rapid transformation of societies*. **2022**. https://www.unep.org/emissions-gap-report-2022
4. Lawrence, M.G.; Schäfer, S.; Muri, H.; Scott, V.; Oschlies, A.; Vaughan, N.E.; Boucher, O.; Schmidt, H.; Haywood, J.; Scheffran, J. Evaluating climate geoengineering proposals in the context of the Paris Agreement temperature goals. *Nat Commun* **2018**, *9*, 3734. https://doi.org/10.1038/s41467-018-05938-3
5. Intergovernmental Panel on Climate Change. *Climate Change 2022: Mitigation of Climate Change. Contribution of Working Group III to the Sixth Assessment Report of the Intergovernmental Panel on Climate Change*; P.R. Shukla, J. Skea, R. Slade, A. Al Khourdajie, R. van Diemen, D. McCollum, M. Pathak, S. Some, P. Vyas, R. Fradera, M. Belkacemi, A. Hasija, G. Lisboa, S. Luz, J. Malley, Eds.; Cambridge University Press, **2022**, doi: 10.1017/9781009157926
6. Schuiling, R.D.; Krijgsman, P. Enhanced Weathering: An Effective and Cheap Tool to Sequester Co2. *Climatic Change*, **2006**, *74*, 349–354. https://doi.org/10.1007/s10584-005-3485-y
7. Köhler, P.; Hartmann, J.; Wolf-Gladrow, D.A. Geoengineering potential of artificially enhanced silicate weathering of olivine. *Proc National Acad Sci*, **2010**, *107*, 20228–20233. https://doi.org/10.1073/pnas.1000545107
8. Renforth, P. The potential of enhanced weathering in the UK. *Int J Greenh Gas Con* **2012**, *10*, 229–243. https://doi.org/10.1016/j.ijggc.2012.06.011
9. Hartmann, J.; West, A.J.; Renforth, P.; Köhler, P.; Rocha, C.L.D.L.; Wolf-Gladrow, D.A.; Dürr, H.H.; Scheffran, J. Enhanced chemical weathering as a geoengineering strategy to reduce atmospheric carbon dioxide, supply nutrients, and mitigate ocean acidification. *Rev Geophys*, **2013**, *51*, 113–149. https://doi.org/10.1002/rog.20004
10. Kantola, I.B.; Masters, M.D.; Beerling, D.J.; Long, S.P.; DeLucia, E.H. Potential of global croplands and bioenergy crops for climate change mitigation through deployment for enhanced weathering. *Biol Letters*, **2017**, *13*, 20160714. https://doi.org/10.1098/rsbl.2016.0714
11. Taylor, L.L.; Beerling, D.J.; Quegan, S.; Banwart, S.A. Simulating carbon capture by enhanced weathering with croplands: an overview of key processes highlighting areas of future model development. *Biol Letters*, **2017**, *13*, 20160868. https://doi.org/10.1098/rsbl.2016.0868
12. Strefler, J.; Amann, T.; Bauer, N.; Kriegler, E.; Hartmann, J. Potential and costs of carbon dioxide removal by enhanced weathering of rocks. *Environ Res Lett*, **2018**, *13*, 034010. https://doi.org/10.1088/1748-9326/aaa9c4

13. Beerling, D.J.; Leake, J.R.; Long, S.P.; Scholes, J.D.; Ton, J.; Nelson, P.N.; Bird, M.; Kantzas, E.; Taylor, L.L.; Sarkar, B.; Kelland, M.; DeLucia, E.; Kantola, I.; Müller, C.; Rau, G.; Hansen, J. Farming with crops and rocks to address global climate, food and soil security. *Nat Plants*, **2018**, *4*, 138–147. https://doi.org/10.1038/s41477-018-0108-y
14. Beerling, D.J.; Kantzas, E.P.; Lomas, M.R.; Wade, P.; Eufrasio, R.M.; Renforth, P.; Sarkar, B.; Andrews, M.G.; James, R.H.; Pearce, C.R.; Mercure, J.-F.; Pollitt, H.; Holden, P.B.; Edwards, N.R.; Khanna, M.; Koh, L.; Quegan, S.; Pidgeon, N.F.; Janssens, I.A.; Hansen, J.; Banwart, S.A. Potential for large-scale $CO_2$ removal via enhanced rock weathering with croplands. *Nature*, **2020**, *583*, 242–248. https://doi.org/10.1038/s41586-020-2448-9
15. Taylor, L.L.; Quirk, J.; Thorley, R.M.S.; Kharecha, P.A.; Hansen, J.; Ridgwell, A.; Lomas, M.R.; Banwart, S.A.; Beerling, D.J. Enhanced weathering strategies for stabilizing climate and averting ocean acidification. *Nat Clim Change*, **2016**, *6*, 402–406. https://doi.org/10.1038/nclimate2882
16. Edwards, D.P.; Lim, F.; James, R.H.; Pearce, C.R.; Scholes, J.; Freckleton, R.P.; Beerling, D.J. Climate change mitigation: potential benefits and pitfalls of enhanced rock weathering in tropical agriculture. *Biol Letters*, **2017**, *13*, 20160715. https://doi.org/10.1098/rsbl.2016.0715
17. Goll, D.S.; Ciais, P.; Amann, T.; Buermann, W.; Chang, J.; Eker, S.; Hartmann, J.; Janssens, I.; Li, W.; Obersteiner, M.; Penuelas, J.; Tanaka, K.; Vicca, S. Potential $CO_2$ removal from enhanced weathering by ecosystem responses to powdered rock. *Nat Geosci,* **2021**, *14*, 545–549. https://doi.org/10.1038/s41561-021-00798-x
18. Rinder, T.; von Hagke, C. The influence of particle size on the potential of enhanced basalt weathering for carbon dioxide removal - Insights from a regional assessment. *J Clean Prod*, **2021**, *315*, 128178. https://doi.org/10.1016/j.jclepro.2021.128178
19. Kantzas, E.P.; Martin, M.V.; Lomas, M.R.; Eufrasio, R.M.; Renforth, P.; Lewis, A.L.; Taylor, L.L.; Mecure, J.-F.; Pollitt, H.; Vercoulen, P.V.; Vakilifard, N.; Holden, P.B.; Edwards, N.R.; Koh, L.; Pidgeon, N.F.; Banwart, S.A.; Beerling, D.J. Substantial carbon drawdown potential from enhanced rock weathering in the United Kingdom. *Nat Geosci*, **2022**, *15*, 382–389. https://doi.org/10.1038/s41561-022-00925-2
20. Zhang, S.; Planavsky, N.J.; Katchinoff, J.; Raymond, P.A.; Kanzaki, Y.; Reershemius, T.; Reinhard, C.T. River chemistry constraints on the carbon capture potential of surficial enhanced rock weathering. *Limnol Oceanogr*, **2022**, https://doi.org/10.1002/lno.12244
21. Haque, F.; Khalidy, R.; Chiang, Y. W.; Santos, R. M. Constraining the Capacity of Global Croplands to $CO_2$ Drawdown via Mineral Weathering. *ACS Earth Space Chem.* **2023**. https://doi.org/10.1021/acsearthspacechem.2c00374.
22. Middelburg, J. J.; Soetaert, K.; Hagens, M. Ocean Alkalinity, Buffering and Biogeochemical Processes. *Reviews of Geophysics,* **2020**., *58*(3), e2019RG000681. https://doi.org/10.1029/2019rg000681
23. Urey, H. C. On the Early Chemical History of the Earth and the Origin of Life. *Proceedings of the National Academy of Sciences,* **1952**, *38*(4), 351–363. https://doi.org/10.1073/pnas.38.4.351
24. Berner, R. A.; Lasaga, A. C.; Garrels, R. M. The carbonate-silicate geochemical cycle and its effect on atmospheric carbon dioxide over the past 100 million years. *American Journal of Science*, **1983**, *283*(7), 641–683. https://doi.org/10.2475/ajs.283.7.641

25. Isson, T. T.; Planavsky, N. J.; Coogan, L. A.; Stewart, E. M.; Ague, J. J.; Bolton, E. W.; Zhang, S.; McKenzie, N. R.; Kump, L. R. Evolution of the Global Carbon Cycle and Climate Regulation on Earth. *Global Biogeochemical Cycles,* **2020**, *34*(2). https://doi.org/10.1029/2018gb006061
26. Smith, P. Soil carbon sequestration and biochar as negative emission technologies. *Global Change Biol,* **2016**, *22*, 1315–1324. https://doi.org/10.1111/gcb.13178
27. Schlesinger, W.H.; Amundson, R. Managing for soil carbon sequestration: Let's get realistic. *Global Change Biol*, **2019**, *25*, 386–389. https://doi.org/10.1111/gcb.14478
28. Bossio, D.A.; Cook-Patton, S.C.; Ellis, P.W.; Fargione, J.; Sanderman, J.; Smith, P.; Wood, S.; Zomer, R.J.; von Unger, M.; Emmer, I.M.; Griscom, B.W. The role of soil carbon in natural climate solutions. *Nat Sustain*, **2020**, *3*, 391–398. https://doi.org/10.1038/s41893-020-0491-z
29. ten Berge, H.F.M.; van der Meer, H.G.; Steenhuizen, J.W.; Goedhart, P.W.; Knops, P.; Verhagen, J. Olivine Weathering in Soil, and Its Effects on Growth and Nutrient Uptake in Ryegrass (Lolium perenne L.): A Pot Experiment. *Plos One*, **2012**, *7*, e42098. https://doi.org/10.1371/journal.pone.0042098
30. Song, Z.; Liu, C.; Müller, K.; Yang, X.; Wu, Y.; Wang, H. Silicon regulation of soil organic carbon stabilization and its potential to mitigate climate change. *Earth-sci Rev* **2018**, *185*, 463–475. https://doi.org/10.1016/j.earscirev.2018.06.020
31. Amann, T.; Hartmann, J.; Struyf, E.; Garcia, W. de O.; Fischer, E.K.; Janssens, I.; Meire, P.; Schoelynck, J. Enhanced Weathering and related element fluxes – a cropland mesocosm approach. *Biogeosciences*, **2020**, *17*, 103–119. https://doi.org/10.5194/bg-17-103-2020
32. Garcia, W. de O.; Amann, T.; Hartmann, J.; Karstens, K.; Popp, A.; Boysen, L.R.; Smith, P.; Goll, D. Impacts of enhanced weathering on biomass production for negative emission technologies and soil hydrology. *Biogeosciences,* **2020**, *17*, 2107–2133. https://doi.org/10.5194/bg-17-2107-2020
33. Kelland, M.E.; Wade, P.W.; Lewis, A.L.; Taylor, L.L.; Sarkar, B.; Andrews, M.G.; Lomas, M.R.; Cotton, T.E.A.; Kemp, S.J.; James, R.H.; Pearce, C.R.; Hartley, S.E.; Hodson, M.E.; Leake, J.R.; Banwart, S.A.; Beerling, D.J. Increased yield and CO2 sequestration potential with the C4 cereal Sorghum bicolor cultivated in basaltic rock dust-amended agricultural soil. *Global Change Biol*, **2020**, *26*, 3658–3676. https://doi.org/10.1111/gcb.15089
34. Swoboda, P.; Döring, T.F.; Hamer, M. Remineralizing soils? The agricultural usage of silicate rock powders: A review. *Sci Total Environ*, **2021**, *807*, 150976. https://doi.org/10.1016/j.scitotenv.2021.150976
35. Guo, F.; Wang, Y.; Zhu, H.; Zhang, C.; Sun, H.; Fang, Z.; Yang, J.; Zhang, L.; Mu, Y.; Man, Y. B.; Wu, F. Crop Productivity and Soil Inorganic Carbon Change Mediated by Enhanced Rock Weathering in Farmland: A Comparative Field Analysis of Multi-Agroclimatic Regions in Central China. *Agric. Syst.* **2023**, *210*, 103691. https://doi.org/10.1016/j.agsy.2023.103691.
36. Luchese, A. V.; Leite, I. J. G. de C.; Alves, M. L.; Vieceli, J. P. dos S.; Pivetta, L. A.; Missio, R. F. Can Basalt Rock Powder Be Used as an Alternative Nutrient Source for Soybeans and Corn? *J. Soil Sci. Plant Nutr.* **2023**, 1–11. https://doi.org/10.1007/s42729-023-01322-3.

37. Reynaert, S.; Vienne, A.; Boeck, H. J. D.; D'Hose, T.; Janssens, I.; Nijs, I.; Portillo-Estrada, M.; Verbruggen, E.; Vicca, S.; Poblador, S. Basalt Addition Improves the Performance of Young Grassland Monocultures under More Persistent Weather Featuring Longer Dry and Wet Spells. *Agric. For. Meteorol.* **2023**, *340*, 109610. https://doi.org/10.1016/j.agrformet.2023.109610.
38. Davies, B.; Finney, B.; Eagle, D. *Resource Management: Soil*. Farming Press, **2001**, ISBN 0 85236 559 4.
39. Kamprath, E.J.; Smyth, T.J. Liming. In: *Encyclopedia of Soils in the Environment*; Hillel, D. Ed;. Elsevier, **2005** ISBN 9780123485304, 350-358, https://doi.org/10.1016/B0-12-348530-4/00225-3
40. Goulding, K.W.T. Soil acidification and the importance of liming agricultural soils with particular reference to the United Kingdom. *Soil Use Manage*, **2016**, *32*: 390-399. https://doi.org/10.1111/sum.12270
41. Semhi, K.; Suchet, P.A.; Clauer, N.; Probst, J.-L. Impact of nitrogen fertilizers on the natural weathering-erosion processes and fluvial transport in the Garonne basin. *Appl Geochem*, **2000**, *15*, 865–878. https://doi.org/10.1016/s0883-2927(99)00076-1
42. West, T.O.; McBride, A.C. The contribution of agricultural lime to carbon dioxide emissions in the United States: dissolution, transport, and net emissions. *Agric Ecosyst Environ*, **2005**, *108*, 145–154. https://doi.org/10.1016/j.agee.2005.01.002
43. Oh, N.; Raymond, P.A. Contribution of agricultural liming to riverine bicarbonate export and CO2 sequestration in the Ohio River basin. *Global Biogeochem Cy*, **2006**, *20*, n/a-n/a. https://doi.org/10.1029/2005gb002565
44. Hamilton, S.K.; Kurzman, A.L.; Arango, C.; Jin, L.; Robertson, G.P. Evidence for carbon sequestration by agricultural liming. *Global Biogeochem Cy*, **2007**, *21*, n/a-n/a. https://doi.org/10.1029/2006gb002738
45. Perrin, A.-S.; Probst, A.; Probst, J.-L. Impact of nitrogenous fertilizers on carbonate dissolution in small agricultural catchments: Implications for weathering CO2 uptake at regional and global scales. *Geochim Cosmochim Ac*, **2008**, *72*, 3105–3123. https://doi.org/10.1016/j.gca.2008.04.011
46. Dietzen, C.; Harrison, R.; Michelsen-Correa, S. Effectiveness of enhanced mineral weathering as a carbon sequestration tool and alternative to agricultural lime: An incubation experiment. *Int J Greenh Gas Con*, **2018**, *74*, 251–258. https://doi.org/10.1016/j.ijggc.2018.05.007
47. Kanzaki, Y.; Zhang, S.; Planavsky, N.J.; Reinhard, C.T. Soil Cycles of Elements simulator for Predicting TERrestrial regulation of greenhouse gases: SCEPTER v0.9. *Geoscientific Model Dev Discuss*, **2022**, 1–58. https://doi.org/10.5194/gmd-2022-8
48. Deng, H.; Sonnenthal, E.; Arora, B.; Breunig, H.; Brodie, E.; Kleber, M.; Spycher, N.; Nico, P. The Environmental Controls on Efficiency of Enhanced Rock Weathering in Soils. *Sci. Rep.* **2023**, *13* (1), 9765. https://doi.org/10.1038/s41598-023-36113-4.
49. Cipolla, G.; Calabrese, S.; Noto, L.V.; Porporato, A. The role of hydrology on enhanced weathering for carbon sequestration I. Modeling rock-dissolution reactions coupled to plant, soil moisture, and carbon dynamics. *Adv Water Resour*, **2021**, *154*, 103934. https://doi.org/10.1016/j.advwatres.2021.103934
50. Renforth, P.; Pogge von Strandmann, P.A.E.; Henderson, G.M. The dissolution of olivine added to soil: Implications for enhanced weathering. *Appl Geochem*, **2015**, *61*, 109–118. https://doi.org/10.1016/j.apgeochem.2015.05.016

51. Pogge von Strandmann, P.A.E; Fraser, W.T.; Hammond, S.J.; Tarbuck, G.; Wood, I.G.; Oelkers, E.H.; Murphy, M.J. Experimental determination of Li isotope behaviour during basalt weathering. *Chem Geol*, **2019**, *517*, 34–43. https://doi.org/10.1016/j.chemgeo.2019.04.020
52. Pogge von Strandmann, P.A.E.; Renforth, P.; West, A.J.; Murphy, M.J.; Luu, T.-H.; Henderson, G.M. The lithium and magnesium isotope signature of olivine dissolution in soil experiments. *Chem Geol*, **2021**, *560*, 120008. https://doi.org/10.1016/j.chemgeo.2020.120008
53. Pogge von Strandmann, P.A.E; Tooley, C.; Mulders, J.J.P.A.; Renforth, P. The Dissolution of Olivine Added to Soil at 4°C: Implications for Enhanced Weathering in Cold Regions. *Frontiers Clim*, **2022**, *4*. https://doi.org/10.3389/fclim.2022.827698
54. Amann, T.; Hartmann, J.; Hellmann, R.; Pedrosa, E.T.; Malik, A. Enhanced weathering potentials—the role of in situ CO2 and grain size distribution. *Frontiers Clim*, **2022**, *4*, 929268. https://doi.org/10.3389/fclim.2022.929268
55. Jariwala, H.; Haque, F.; Vanderburgt, S.; Santos, R.M.; Chiang, Y.W. Mineral–Soil–Plant–Nutrient Synergisms of Enhanced Weathering for Agriculture: Short-Term Investigations Using Fast-Weathering Wollastonite Skarn. *Front Plant Sci*, **2022**, *13*, 929457. https://doi.org/10.3389/fpls.2022.929457
56. Vienne, A.; Poblador, S.; Portillo-Estrada, M.; Hartmann, J.; Ijiehon, S.; Wade, P.; Vicca, S. Enhanced Weathering Using Basalt Rock Powder: Carbon Sequestration, Co-benefits and Risks in a Mesocosm Study With Solanum tuberosum. *Frontiers Clim*, **2022**, *4*, 869456. https://doi.org/10.3389/fclim.2022.869456
57. Haque, F.; Santos, R.M.; Chiang, Y.W. CO2 sequestration by wollastonite-amended agricultural soils – An Ontario field study. *Int J Greenh Gas Con*, **2020**, *97*, 103017. https://doi.org/10.1016/j.ijggc.2020.103017
58. Taylor, L.L.; Driscoll, C.T.; Groffman, P.M.; Rau, G.H.; Blum, J.D.; Beerling, D.J. Increased carbon capture by a silicate-treated forested watershed affected by acid deposition. *Biogeoscience*, **2021**, *18*, 169–188. https://doi.org/10.5194/bg-18-169-2021
59. Larkin, C.S.; Andrews, M.G.; Pearce, C.R.; Yeong, K.L.; Beerling, D.J.; Bellamy, J.; Benedick, S.; Freckleton, R.P.; Goring-Harford, H.; Sadekar, S.; James, R.H. Quantification of CO2 removal in a large-scale enhanced weathering field trial on an oil palm plantation in Sabah, Malaysia. *Frontiers Clim*, **2022**, *4*, 959229. https://doi.org/10.3389/fclim.2022.959229
60. Dietzen, C.; Rosing, M. T. Quantification of CO2 Uptake by Enhanced Weathering of Silicate Minerals Applied to Acidic Soils. *Int J Greenh Gas Con* **2023**, *125*, 103872. https://doi.org/10.1016/j.ijggc.2023.103872.
61. Calabrese, S.; Wild, B.; Bertagni, M.B.; Bourg, I.C.; White, C.; Aburto, F.; Cipolla, G.; Noto, L.V.; Porporato, A. Nano- to Global-Scale Uncertainties in Terrestrial Enhanced Weathering. *Environ Sci Technol*, **2022**, https://doi.org/10.1021/acs.est.2c03163
62. Almaraz, M.; Bingham, N.L.; Holzer, I.O.; Geoghegan, E.K.; Goertzen, H.; Sohng, J.; Houlton, B.Z. Methods for determining the CO2 removal capacity of enhanced weathering in agronomic settings. *Frontiers Clim*, **2022**, *4*, 970429. https://doi.org/10.3389/fclim.2022.970429
63. Chay, F.; Klitzke, J.; Hausfather, Z.; Martin, K.; Freeman, J.; Cullenward, D. 2022. Verification Confidence Levels for carbon dioxide removal. CarbonPlan https://carbonplan.org/research/cdr-verification-explainer (accessed 2023-05-08).

64. Knapp, W. J.; Stevenson, E. I.; Renforth, P.; Ascough, P. L.; Knight, A. C. G.; Bridgestock, L.; Bickle, M. J.; Lin, Y.; Riley, A. L.; Mayes, W. M.; Tipper, E. T. Quantifying CO2 Removal at Enhanced Weathering Sites: A Multiproxy Approach. *Environ. Sci. Technol.* **2023**, *57* (26), 9854–9864. https://doi.org/10.1021/acs.est.3c03757.
65. Amann, T.; Hartmann, J. Carbon Accounting for Enhanced Weathering. *Frontiers Clim* **2022**, *4*, 849948. https://doi.org/10.3389/fclim.2022.849948.
66. Corwin, D. L.; Lesch, S. M. Apparent Soil Electrical Conductivity Measurements in Agriculture. *Comput Electron Agr* **2005**, *46* (1–3), 11–43. https://doi.org/10.1016/j.compag.2004.10.005.
67. Brimhall, G.H.; Dietrich, W.E. Constitutive mass balance relations between chemical composition, volume, density, porosity, and strain in metasomatic hydrochemical systems: Results on weathering and pedogenesis. *Geochim Cosmochim Ac*, **1987**, *51*, 567–587. https://doi.org/10.1016/0016-7037(87)90070-6
68. Chadwick, O.A.; Brimhall, G.H.; Hendricks, D.M. From a black to a gray box — a mass balance interpretation of pedogenesis. *Geomorphology*, **1990**, *3*, 369–390. https://doi.org/10.1016/0169-555x(90)90012-f
69. Chadwick, O.A.; Derry, L.A.; Vitousek, P.M.; Huebert, B.J.; Hedin, L.O. Changing sources of nutrients during four million years of ecosystem development. *Nature*, **1999**, *397*, 491–497. https://doi.org/10.1038/17276
70. Brimhall, G.H.; Lewis, C.J.; Ford, C.; Bratt, J.; Taylor, G.; Warin, O. Quantitative geochemical approach to pedogenesis: importance of parent material reduction, volumetric expansion, and eolian influx in lateritization. *Geoderma*, **1991**, 51, 51–91. https://doi.org/10.1016/0016-7061(91)90066-3
71. Kurtz, A.C.; Derry, L.A.; Chadwick, O.A.; Alfano, M.J. Refractory element mobility in volcanic soils. *Geology*, **2000**, *28*, 683–686. https://doi.org/10.1130/0091-7613(2000)28<683:remivs>2.0.co;2
72. White, A.F.; Bullen, T.D.; Schulz, M.S.; Blum, A.E.; Huntington, T.G.; Peters, N.E. Differential rates of feldspar weathering in granitic regoliths. *Geochim Cosmochim Ac* **2001**, *65*, 847–869. https://doi.org/10.1016/s0016-7037(00)00577-9
73. Anderson, S.P.; Dietrich, W.E.; Brimhall, G.H. Weathering profiles, mass-balance analysis, and rates of solute loss: Linkages between weathering and erosion in a small, steep catchment. *GSA Bulletin*, **2002**, *114*, 1143–1158. https://doi.org/10.1130/0016-7606(2002)114<1143:wpmbaa>2.0.co;2
74. Riebe, C.S.; Kirchner, J.W.; Finkel, R.C. Long-term rates of chemical weathering and physical erosion from cosmogenic nuclides and geochemical mass balance. *Geochim Cosmochim Ac*, **2003**, *67*, 4411–4427. https://doi.org/10.1016/s0016-7037(03)00382-x
75. Tabor, N.J.; Montañez, I.P.; Zierenberg, R.; Currie, B.S. Mineralogical and geochemical evolution of a basalt-hosted fossil soil (Late Triassic, Ischigualasto Formation, northwest Argentina): Potential for paleoenvironmental reconstruction. *GSA Bulletin*, **2004**, *116*, 1280–1293. https://doi.org/10.1130/b25222.1
76. Sheldon, N.D.; Tabor, N.J. Quantitative paleoenvironmental and paleoclimatic reconstruction using paleosols. *Earth-sci Rev*, **2009**, *95*, 1–52. https://doi.org/10.1016/j.earscirev.2009.03.004
77. Brantley, S.L.; Lebedeva, M. Learning to Read the Chemistry of Regolith to Understand the Critical Zone. *Annu Rev Earth Pl Sc*, **2011**, *39*, 387–416. https://doi.org/10.1146/annurev-earth-040809-152321

78. Fisher, B.A.; Rendahl, A.K.; Aufdenkampe, A.K.; Yoo, K. Quantifying weathering on variable rocks, an extension of geochemical mass balance: Critical zone and landscape evolution. *Earth Surf. Process. Landforms*, **2017**, *42*, 2457–2468. https://doi.org/10.1002/esp.4212
79. Lipp, A.G.; Shorttle, O.; Sperling, E.A.; Brocks, J.J.; Cole, D.B.; Crockford, P.W.; Mouro, L.D.; Dewing, K.; Dornbos, S.Q.; Emmings, J.F.; Farrell, U.C.; Jarrett, A.; Johnson, B.W.; Kabanov, P.; Keller, C.B.; Kunzmann, M.; Miller, A.J.; Mills, N.T.; O'Connell, B.; Peters, S.E.; Planavsky, N.J.; Ritzer, S.R.; Schoepfer, S.D.; Wilby, P.R.; Yang, J. The composition and weathering of the continents over geologic time. *Geochem Perspectives Lett*, **2021**, 21–26. https://doi.org/10.7185/geochemlet.2109
80. Middelburg, J.J.; van der Weijden, C.H.; Woittiez, J.R.W. Chemical processes affecting the mobility of major, minor and trace elements during weathering of granitic rocks. *Chem Geol*, **1988**, 68, 253–273. https://doi.org/10.1016/0009-2541(88)90025-3
81. Pennock, D.; Yates, T.; Braidek, J. Soil Sampling Designs. In: *Soil Sampling and Methods of Analysis*. Carter, M.R.; Gregorich, E.G.; eds. CRC Press, Canadian Society of Soil Science, **2008**, 23-38.
82. Knapp, W. J.; Tipper, E. T. The Efficacy of Enhancing Carbonate Weathering for Carbon Dioxide Sequestration. *Frontiers Clim* **2022**, *4*, 928215. https://doi.org/10.3389/fclim.2022.928215.
83. Harrington, K. J.; Hilton, R. G.; Henderson, G. M. Implications of the Riverine Response to Enhanced Weathering for CO2 Removal in the UK. *Appl Geochem* **2023**, *152*, 105643. https://doi.org/10.1016/j.apgeochem.2023.105643.
84. Kanzaki, Y.; Planavsky, N. J.; Reinhard, C. T. New Estimates of the Storage Permanence and Ocean Co-Benefits of Enhanced Rock Weathering. *PNAS Nexus* **2023**, *2* (4), pgad059. https://doi.org/10.1093/pnasnexus/pgad059.
85. Rousseau, R. M. Detection Limit and Estimate of Uncertainty of Analytical XRF Results. *Rigaku J*. **2001**, *18* (2), 33-47.
86. Krishna, A. K.; Murthy, N. N.; Govil, P. K. Multielement Analysis of Soils by Wavelength-Dispersive X-Ray Fluorescence Spectrometry. *Atom. Spectrosc*. **2007**, *28* (6), 202-214.
87. Kenna, T. C.; Nitsche, F. O.; Herron, M. M.; Mailloux, B. J.; Peteet, D.; Sritrairat, S.; Sands, E.; Baumgarten, J. Evaluation and Calibration of a Field Portable X-Ray Fluorescence Spectrometer for Quantitative Analysis of Siliciclastic Soils and Sediments. *J. Anal. At. Spectrom.* **2010**, *26* (2), 395–405. https://doi.org/10.1039/c0ja00133c.
88. Andersen, J.E.T. On the development of quality assurance. *Trac Trends Anal Chem*, **2014**, *60*, 16–24. https://doi.org/10.1016/j.trac.2014.04.016
89. Eggen, O.A.; Reimann, C.; Flem, B. Reliability of geochemical analyses: Deja vu all over again. *Sci Total Environ*, **2019**, *670*, 138–148. https://doi.org/10.1016/j.scitotenv.2019.03.185
90. Inghram, M.G. Stable isotope dilution as an analytical tool. *Ann Rev Nucl Sci*, **1954**, *4:*1, 81-92. https://doi.org/10.1146/annurev.ns.04.120154.000501
91. Evans, E.H.; Clough, R. Isotope dilution analysis. In: *Encyclopedia of Analytical Science,* 2nd ed; Worsfold, P.; Townshend, A.; Poole., C.; Eds.. Elsevier, **2005**, 545-553. https://doi.org/10.1016/B0-12-369397-7/00301-0

92. Stracke, A.; Scherer, E.E.; Reynolds, B.C. Application of isotope dilution in geochemistry. *Treatise on Geochemistry*, 2nd ed., **2014**, 71–86. https://doi.org/10.1016/b978-0-08-095975-7.01404-2
93. Willbold, M.; Jochum, K.P. Multi-Element Isotope Dilution Sector Field ICP-MS: A Precise Technique for the Analysis of Geological Materials and its Application to Geological Reference Materials. *Geostand Geoanal Res*, **2005**, *29*, 63–82. https://doi.org/10.1111/j.1751-908x.2005.tb00656.x
94. Raymond, P.A.; Saiers, J.E.; Sobczak, W.V. Hydrological and biogeochemical controls on watershed dissolved organic matter transport: pulse-shunt concept. *Ecology*, **2016**, *97*, 5–16. https://doi.org/10.1890/14-1684.1
95. Siever, R.; Woodford, N. Dissolution kinetics and the weathering of mafic minerals. *Geochim Cosmochim Ac*, **1979**, *43*, 717–724. https://doi.org/10.1016/0016-7037(79)90255-2
96. Velbel, M.A. Influence of surface area, surface characteristics, and solution composition on feldspar weathering rates. In: *Geochemical Processes at Mineral Surfaces* J.A. Davis; K.F. Hayes; Eds. American Chemical Society Symposium Series No. 323, **1986**, 615-634
97. White, A.F.; Peterson, M.L. Chemical Modeling of Aqueous Systems II. *Acs Sym Ser*, **1990**, 461–475. https://doi.org/10.1021/bk-1990-0416.ch035
98. Appelo, C. A. J. Multicomponent Ion Exchange and Chromatography in Natural Systems. *Reviews in Mineralogy and Geochemistry 1* (34), 193–227.
99. Kahouli-Brahmi, S. Technological learning in energy–environment–economy modelling: A survey. *Energ Policy*, **2008**, *36*, 138–162. https://doi.org/10.1016/j.enpol.2007.09.001
100. Nugent, M.A.; Brantley, S.L.; Pantano, C.G.; Maurice, P.A. The influence of natural mineral coatings on feldspar weathering. *Nature*, **1998**, *395*, 588–591. https://doi.org/10.1038/26951
101. White, A.F.; Brantley, S.L. The effect of time on the weathering of silicate minerals: why do weathering rates differ in the laboratory and field? *Chem Geol*, **2003**, *202,* 479–506. https://doi.org/10.1016/j.chemgeo.2003.03.001
102. Béarat, H.; McKelvy, M.J.; Chizmeshya, A.V.G.; Gormley, D.; Nunez, R.; Carpenter, R.W.; Squires, K.; Wolf, G.H. Carbon Sequestration via Aqueous Olivine Mineral Carbonation: Role of Passivating Layer Formation. *Environ Sci Technol*, **2006**, *40*, 4802–4808. https://doi.org/10.1021/es0523340
103. Daval, D.; Calvaruso, C.; Guyot, F.; Turpault, M.-P. Time-dependent feldspar dissolution rates resulting from surface passivation: Experimental evidence and geochemical implications. *Earth Planet Sc Lett*, **2018**, *498*, 226–236. https://doi.org/10.1016/j.epsl.2018.06.035
104. Zhu, C.; Lu, P. Alkali feldspar dissolution and secondary mineral precipitation in batch systems: 3. Saturation states of product minerals and reaction paths. *Geochim Cosmochim Ac*, **2009**, *73*, 3171–3200. https://doi.org/10.1016/j.gca.2009.03.015
105. Emmanuel, S.; Ague, J.J. Impact of nano-size weathering products on the dissolution rates of primary minerals. *Chem Geol*, **2011**, *282*, 11–18. https://doi.org/10.1016/j.chemgeo.2011.01.002
106. White, A.F.; Schulz, M.S.; Lawrence, C.R.; Vivit, D.V.; Stonestrom, D.A. Long-term flow-through column experiments and their relevance to natural granitoid

weathering rates. *Geochim Cosmochim Ac*, **2017**, *202*, 190–214. https://doi.org/10.1016/j.gca.2016.11.042
107. Brantley, S.L.; Shaughnessy, A.; Lebedeva, M.I.; Balashov, V.N. How temperature-dependent silicate weathering acts as Earth's geological thermostat. *Science*, **2023**, *379*, 382–389. https://doi.org/10.1126/science.add2922
108. Austin, R.; Gatiboni, L.; Havlin, J. Soil Sampling Strategies for Site-Specific Field Management. NC State Extensions, 2020, https://content.ces.ncsu.edu/soil-sampling-strategies-for-site-specific-field-management (accessed 2023-05-08).
109. Allen, W.J.; Sapsford, S.J.; Dickie, I.A. Soil sample pooling generates no consistent inference bias: a meta-analysis of 71 plant–soil feedback experiments. *New Phytol*, **2021**, *231*, 1308–1315. https://doi.org/10.1111/nph.17455
110. Carroll, D. Ion exchange in clays and other minerals. *GSA Bulletin*, **1959**, *70*, 749–779. https://doi.org/10.1130/0016-7606(1959)70[749:ieicao]2.0.co;2
111. Sumner, M.E.; Miller, W.P. Cation Exchange Capacity and Exchange Coefficients. In: *Methods of Soil Analysis Part 3: Chemical Methods*. Sparks, D.L.; ed. SSSA Book Series 5, Soil Science Society of America, **1996**, 1201-1230.
112. Sparks, D.L. *Environmental soil chemistry*, 2$^{nd}$ ed. Academic Press, 2003., ISBN 9780126564464, https://doi.org/10.1016/B978-012656446-4/50006-2.

# Supporting Information

## Initial validation of a soil-based mass-balance approach for empirical monitoring of enhanced rock weathering rates

Tom Reershemius[1*] and Mike E. Kelland[2*], Jacob S. Jordan[3], Isabelle R. Davis[1,4], Rocco D'Ascanio[1], Boriana Kalderon-Asael[1], Dan Asael[1], T. Jesper Suhrhoff[5,1], Dimitar Z. Epihov[2], David J. Beerling[2], Christopher T. Reinhard[6], Noah J. Planavsky[1,5]

[1]Department of Earth and Planetary Sciences, Yale University, New Haven, CT, USA
[2]Leverhulme Centre for Climate Change Mitigation, School of Biosciences, University of Sheffield, Sheffield, UK
[3]Porecast Research, Lawrence, KS, USA
[4]School of Ocean and Earth Science, University of Southampton Waterfront Campus, Southampton, UK
[5]Yale Center for Natural Carbon Capture, Yale University, New Haven, CT, USA
[6]School of Earth and Atmospheric Sciences, Georgia Institute of Technology, GA, USA

**\*corresponding and equal contribution authors:** tom.reershemius@yale.edu, mekelland1@sheffield.ac.uk

**1 Supporting Materials and Methods** *(ppS1-S10)*
**1.1** Mesocosm design and construction
**1.2** Substrate preparation and characterization
**1.3** Basaltic rock preparation and characterization
**1.4** Plant varieties and growth conditions
**1.5** Irrigation regime
**1.6** Leachate, soil exchangeable fraction and plant sample preparation and analysis
**1.7** Soil sample preparation and analysis
**1.8** Calculating ERW and CDR rates from leachate, soil exchangeable fraction and plant samples
**1.9** Calculating ERW and CDR rates using the TiCAT theoretical scheme
**1.10** Correction for weathering due to strong acids

**2 Supporting Tables** *(ppS11-S12)*
**Table S1**: Chemical composition (major oxides) of basalt feedstock used, Columbia River Basalt, Prineville Chemical Type Unit
**Table S2**: Chemical composition of artificial rainwater irrigation solution

**3 Supporting Figures** *(ppS13-S22)*
**Figure S1**: Mesocosm design
**Figure S2**: Cumulative particle size distribution of basalt feedstock
**Figure S3**: Mineralogy of basalt feedstock
**Figure S4:** Irrigation schedule
**Figure S5**: Isotopic composition of sample and spike solutions for Mg, Ti, and Ca
**Figure S6**: The TiCAT conceptual framework as a three-component "mixing triangle"
**Figure S7**: Concentration in ashed solid samples of Mg, Ca, and Na from the "topsoil" depth interval (0-12 cm) in all basalt-amended mesocosms in this study
**Figure S8**: Depth profiles of concentration in ashed solid samples of Ti, Ca, Mg, and Na from mesocosms in this study
**Figure S9**: Depth profiles of ratios in concentration in ashed solid samples of (Ca+Mg+Na)/Ti, Ca/Ti, Mg/Ti, and Na/Ti from mesocosms in this study
**Figure S10**: Box and whisker plot showing % recovery of basalt feedstock in samples from all basalt-treated mesocosms

**4 Supporting References** *(ppS23-S26)*

**5 Supporting Datasets** *(see attached files)*
**Dataset S1:** All raw data (with errors) for leachate, plant, soil exchangeable fraction, soil solid phase
**Dataset S2:** attached file "mesocosms_initial_soil_20230219.xlsx", required to run Matlab script "TiCAT_X"
**Dataset S3:** attached file "feedstocks.xlsx", required to run Matlab script "TiCAT_X"
**Software S1:** attached file "TiCAT_X.doc", Matlab script in text format

# 1 Supporting Materials and Methods

## 1.1 Mesocosm design and construction

Replicated mesocosms (**Fig. S1a**) ($n$ = 7) for the four soil treatments in this study (NPK –basalt, NPK +basalt, manure –basalt, manure +basalt) were constructed from two sections of polyvinyl chloride (PVC) pipe, joined by a pair of PVC endcaps fixed in a back-to-back arrangement (152 mm internal diameter pipe, 160 mm internal diameter endcaps; Brett Martin, Newtownabbey, UK). The freestanding mesocosms were stabilised with a 170 mm × 190 mm acrylic baseplate fixed to a further 160 mm internal diameter endcap that was attached to the lower pipe section. Weight-bearing external joints were secured with PVC weld cement (Tangit ABS; Henkel, Düsseldorf, Germany) and internal seals were formed from high-strength polyurethane sealant (PU18; Bond It Ltd, Elland, UK).

Soil was loaded into the 600 mm long top section and leachate collection apparatus were housed in the 400 mm long lower section. The joint attaching these pipe sections was fitted with a 20 mm internal diameter nylon nozzle to direct leachate into a high-density polyethylene bottle via a polypropylene funnel (1 L narrow-neck polyethylene bottle, 150 mm lightweight polypropylene filter funnel; Azlon, Stoke-on-Trent, UK). Soil particle migration into the leachate nozzle was reduced by emplacing a 60 mm deep drainage layer in the base of the substrate-holding top section, comprised of 20 mm diameter polyoxymethylene balls (Bearing Warehouse Ltd, Sheffield, UK). A detachable 3-ply opaque mylar membrane (Reflex Total Blackout; GroWell Horticulture Ltd, Sheffield, UK) was used to reduce algal growth in the leachate collection apparatus.

## 1.2 Substrate preparation and characterization

Mildly acidic (pH 6.45) agricultural soil of clay-loam texture (29.0, 34.3 and 36.7% by mass of clay (<2 µm), silt (2-60 µm) and sand (60-2,000 µm), respectively) was collected from farmland under *Brassica napus* (oil seed rape) cultivation at the Game and Wildlife Conservation Trust, Allerton Project, Leicestershire, UK (latitude 52.611286 ºN, longitude 0.831559 ºW). The soil had an initial cation exchange capacity of 22.0 cmol ($p^+$) kg$^{-1}$, with a base saturation of 70.0%, calcium saturation of 63.4%, and a gravimetric water content (GWC) at field capacity of 0.68 g g$^{-1}$. The soil total carbon concentration was 2.3 wt% and the soil organic carbon concentration was 1.1 wt%.

Soil was collected from a 2 m x 1 m trench, about 0.20 - 0.25 m deep. 400 kg (wet weight) of soil was collected from the field. Collected soil was manually comminuted and homogenized at 18°C and stored in partially sealed polypropylene bags to reduce moisture loss. Large stones and invertebrates were removed before the soil was passed through a woven stainless-steel sieve (2 mm mesh size; Fisher Scientific UK Ltd, Loughborough, UK). After comminution, the soil had a GWC of 0.28 g g$^{-1}$. A split of comminuted, homogenized soil was taken for compositional analysis (see *Supporting Information Section 1.7*). Organic fertilizer for the experiment was derived from cow manure collected from Hope Farm, Hassop, UK. The manure was air dried for 30 days, then manually comminuted at 18 °C and passed through a 2 mm mesh size stainless-steel sieve.

Mesocosms were filled with substrate in a sub- and top-soil arrangement to simulate a tillage regime (3:1 mass ratio of sub- to top-soil) **(Fig. S1b)**. The total dry mass of soil in each mesocosm was 8.252 kg, which – being compacted to a column height of 500 mm – had an average bulk density of 0.91 kg L$^{-1}$. Basalt and fertilizer were added, as required, to the topsoil only and were mixed thoroughly into the substrate immediately prior to mesocosm filling on day 0. A total mass of 91 g (equivalent to 5.0 kg m$^{-2}$) of basalt (see below) was added to treated mesocosms. Soils treated with chemical fertilizer received individually weighed quantities of analytical-grade compounds (0.347 g of urea, 1.633 g of powdered diammonium phosphate and 1.482 g of powdered potassium chloride), which were estimated to compensate for soil nutrient losses to the crop and leachate during the experiment. Mesocosms treated with organic fertilizer received 17.7 g (equivalent to 1.0 kg m$^{-2}$) of air-dried cow manure, which contained 2.57, 1.23 and 3.08% total nitrogen, phosphorus pentoxide and potassium oxide, respectively. The manure application rate was calculated to approximately equal the total nitrogen application rate in the chemical fertilizer treatment, being 25 g N m$^{-2}$ (250 kg N ha$^{-1}$).

Initial soil pH was measured using a Jenway 3540 Bench Combined Conductivity/pH Meter (Jenway, Stone, UK) calibrated with standard solutions at pH 4, 7 and 10. Prior to pH determination, 15 g of air-dried soil was thoroughly mixed with 30 ml of deionized water in a 100 ml borosilicate glass beaker using a glass stirrer. The solution was then left for 30 minutes before recording the pH of the slurry to 0.01 pH units. Soil carbon measurements were made using a Vario EL Cube Carbon-Nitrogen Analyzer (Elementar, Stockport, UK), with soil organic carbon content determined by comparing carbon concentrations before and after soil treatment with 6 M HCl (soil:acid ratio of 90 mg to 0.5 mL).

### 1.3 Basaltic rock preparation and characterization

Crushed basalt of middle Miocene age and belonging to the 'Prineville Chemical Type Unit' of the Columbia River Basalt was sourced from the Cascade Range, Oregon (Central Oregon Basalt Products LLC, Madras, ORE, USA). The basalt was passed through a series of stainless-steel sieves of decreasing mesh size (250 µm, 100 µm and 50 µm mesh size), suspended in analytical-grade ethanol and sonicated for 30 minutes in a 5 L Pyrex beaker to separate submicron particles from larger grains. The suspension was then left for 24 h (sufficient to settle particles with diameters > 1 µm) after which the supernatant was decanted and the settled basalt material was exposed to the air for 72 h to evaporate the residual ethanol. The particle size distribution of the treated basalt was measured using an LA-950 laser diffraction particle size distribution analyzer (Horiba UK Ltd, Northampton, UK), where the particle diameter range was found to be 1.5-300 µm and the $p80$ value (the size at which 80% of the particles have diameters less than or equal to) was 35 µm (**Fig. S2**).

Specific surface area of the treated basalt was determined using an Anton Paar Nova 800 Brunauer Emmett Tell (BET) surface area analyzer in the Yale Geochemistry Center and found to be 10.22 m$^2$ g$^{-1}$. Error was estimated to be 0.01 m$^2$ g$^{-1}$, based on a 5-point analysis. The surface area measured demonstrates the limitations of modelling mineral weathering using a shrinking core model (1):

We first consider a container with a volume 1 cm$^3$, V$_c$. and spheres, V$_s$, given by the $p80$ (35 µm) of the basalt in the study. The maximum number of perfect spheres in this container is given by:

$$n \sim \frac{\pi}{3\sqrt{2}} * \frac{V_c}{V_s} \quad . \quad (S1)$$

Using equation S1 the number of ideally packed spheres with uniform particle diameter of 35 μm is ~ $3.29 \times 10^7$. The surface area of a single sphere with diameter of 35 μm is $3.85 \times 10^{-9}$ m². Therefore, the upper bound of total surface area within a 1 cm³ control volume is 0.127 m² cm⁻³. Assuming ideally packed spheres, the solid phase occupies a volume fraction of $\frac{\pi}{3\sqrt{2}}$ within the control volume. Therefore, assuming density of basalt ~3 g cm⁻³, the specific surface area of ideally packed spheres is 0.0570 m² g⁻¹. Given our assumptions, this is a high upper bound for surface area because we assume close packing, yet consider all surface area available for reaction.

The specific surface area calculated from sphere packing of grains at the diameter of the $p80$ value here is ~$10^4$ m² g⁻¹ smaller than the measured specific surface area. Consequently, the reactive surface area available for dissolution reactions to take place is significantly higher than a model based on a simplified geometry suggests. Nonetheless, the majority of the reactive surface area initially comes from the smallest particles in the size distribution; once weathering ensues, this surface area will likely decrease.

The mineralogy of the basalt (**Fig. S3**) was determined using X-ray diffraction (XRD) analysis at the British Geological Survey (BGS) Keyworth laboratories, following the protocol reported in ref. 2. The basalt contained no detectable carbonate minerals – something which basalt feedstocks should be screened for prior to use as ERW feedstocks, as this reduces a feedstock's carbon dioxide removal potential. The carbon dioxide removal potential of the basalt feedstock, $E_{pot}$ (i.e. the maximum amount of carbon dioxide consumed if all basalt were to react with proton equivalents from carbonic acid, and carbon sequestered as bicarbonate), can be quantified using a modified Steinour formulation, which relates maximum theoretical carbon dioxide reaction with elemental composition of silicate feedstocks (3; see e.g., 4-7). The Steinour formulation using major cations for alkaline minerals is given as:

$$E_{pot} = \frac{\eta \, M_{CO_2}}{100} * \left( \frac{\alpha \, CaO}{M_{CaO}} + \frac{\beta \, MgO}{M_{MgO}} + \frac{\varepsilon \, Na_2O}{M_{Na_2O}} + \frac{\theta \, K_2O}{M_{K_2O}} \right) \quad , \quad (S2)$$

where $M_i$ is molar mass; $\eta$, $\alpha$, $\beta$, $\varepsilon$, and $\theta$ are stoichiometric constants (with values 2, 1, 1, 1, and 1 respectively); and CaO, MgO, Na₂O and K₂O are abundances of major oxides in the feedstock. Using this formulation, with the values for major oxide abundance given in **Table S1**, $E_{pot}$ for the basalt feedstock used here is 211.81 kgCO₂ t⁻¹. This assessment is a conservative estimate representing incongruent dissolution (i.e., does not include cations that will react to form secondary minerals, including $Al^{3+}$ and $Fe^{2+}$; e.g., 8). Note that some previous authors have simplified this formulation to only include CaO and MgO, when assessing $E_{pot}$ for fast-weathering Ca- and Mg-silicates (e.g., 9), which are the most abundant cations. In this study, we ignore the contribution of K⁺ to carbon dioxide removal, due to its presence in NPK fertilizers; as such, when excluding K₂O, $E_{pot}$ for the basalt feedstock is revised to 183.56 kgCO₂ t⁻¹ for the purposes of calculating CDR efficiency in this study (see *Main Text Section 3*).

### 1.4 Plant varieties and growth conditions

A single specimen of dwarf hybrid *Sorghum bicolor* (Pennsylvania 115, Oakbank Game and Conservation, Cambridgeshire, UK) was cultivated in each mesocosm. Approximately 600 seeds were germinated on filter paper (Whatman #1; Cytiva, Sheffield, UK) moistened with distilled water, placed in grip seal polyethylene bags and incubated at 25°C in darkness for 24 h. 144 seeds of similar radicle length (ca. 5 mm) were selected for planting in 24-cell seed trays containing soil from the same field site. Seed trays were transferred to a growth room and maintained at 60–70% relative humidity, with an 18 h day length and 25/17°C day/night temperatures. After 24 h, 87 seedlings emerged from the soil, of which the 28 largest were grown for a further 9 d and chosen at random for transplantation to mesocosms on day 0.

Additive $CO_2$ was used to sustain atmospheric concentrations of 400 ppm in the controlled environment and a photosynthetic photon flux density of 800 µmol photons m$^{-2}$ s$^{-1}$ was maintained throughout the 18 h days between emergence and boot stage (days 0-60). From day 61 onwards the day length was reduced to 10 h and on day 115 (at physiological maturity) *Sorghum* seeds and shoots were harvested. Henceforth a fallow period was simulated by reducing the day/night temperatures to 5/3°C and maintaining relative humidity at 75% until the experiment was terminated on day 235.

### 1.5 Irrigation regime

Throughout the *Sorghum* lifecycle, mesocosms were weighed every 3 days using a portable laboratory balance (Navigator NVT; Ohaus, Parsippany, NJ, USA) to estimate depth-averaged soil gravimetric water content (GWC). An artificial rainwater solution (**Table S2**) was then used to approximately compensate for soil water losses via evapotranspiration, with equal volumes of solution being supplied to each mesocosm (**Fig. S4**). Irrigation water was delivered manually to the top surface of each soil column using a 1.5 L low-pressure portable sprayer (Exo Terra, Castleford, UK), with amounts measured gravimetrically to the nearest gram. Approximately every 12 d between day 0 and day 81, equal volumes of additional irrigation solution were supplied to all soil columns to produce leachate for geochemical analysis. During the simulated fallow period (days 115-235) mesocosms were weighed every 14 d and variably irrigated to maintain soil GWC at 0.38 g g$^{-1}$.

### 1.6 Leachate, soil exchangeable fraction and plant sample preparation and analysis

Leachate was collected at six discrete events throughout the experiment. Leachate was prepared for cation analysis by filtration of subsamples through a 0.45 µm cellulose nitrate filter (Minisart: Sartorius, Goettingen, Germany) and stabilization in a matrix of 2% (vol%) analytical-grade nitric acid. Harvested *Sorghum* root, shoot and seed tissues (collected after 235 days) were dried in a forced-air oven at 60 ºC for 48 h and comminuted separately using a laboratory grinding mill (Model: A10 S2, IKA-Werke GmbH & CO. KG, Staufen, Germany). Oven-dried, powdered tissue samples were then subjected to a 45-min microwave-assisted (Model: Multiwave; Anton Paar Ltd, St Albans, UK) nitric acid-hydrogen peroxide digest (0.2 g tissue:3 mL nitric acid:3 mL MQ water:2ml hydrogen peroxide), after which extracts were prepared for cation analysis by centrifugation and filtration of the supernatant through a 0.45 µm filter. Multi-element cation analysis of all prepared leachate, plant and soil extracts was conducted at the School of Biosciences, University of Nottingham (Sutton Bonington Campus) using an iCAP Q ICP-MS (Thermo Fisher Scientific, Bremen, Germany).

### 1.7 Soil sample preparation and analysis

Soil samples were collected from all mesocosms after 235 days. Substrate was excavated from the mesocosms in five separate layers: the topsoil/amended layer (0-120 mm depth), three intermediate subsoil layers (120-240, 240-360, 360-480 mm depth) and the bottommost subsoil layer (480-500 mm depth). Each soil layer was manually comminuted into a bulk sample, homogenized, air dried at 60 ºC for 48 h, and passed through a 2mm steel sieve. Note that the grain size distribution of the basalt feedstock added (see **Fig. S2**) was such that this sieving still allowed close to 100% of the basalt to pass through the sieve. 100 mg of dried sample was subjected to leaching with a 1N ammonium acetate solution at pH = 8.2 in order to remove the exchangeable fraction from the solid phase (after 10). The resultant solid sample was dried at 60˚C overnight, then heated at 600 ˚C for 24 hours to remove organic matter in the soil mixture. The ashed sample was then dissolved in a mixture of aqua regia (hydrochloric and nitric acid) and hydrofluoric acid as follows: 10ml aqua regia was added to Teflon beakers containing ashed sample, left to sit for 4 hours after which 1ml hydrofluoric acid was added, the beakers capped and left on a hotplate for 24 hours at 100˚C. Samples were then uncapped and left to dry down on a hotplate at 90˚C. Samples were raised in nitric acid and diluted to run on a Thermo Scientific iCAP Triple Quadropole inductively coupled plasma mass spectrometer (ICP-MS) at the Yale Geochemistry Center. In order to reduce analytical error, we measured the concentrations of key elements (Ca, Mg, Ti) using an isotope dilution method (see *Main Text Section 2.2*). It should also be noted that there are other methods available to digest bulk solid soil samples for ICP-MS analysis without the use of HF, such as using lithium tetraborate fusion.

A common practice to measure elemental concentration in a variety of industrial applications is X-ray fluorescence (XRF). XRF is an analytical technique used for measuring the elemental concentration of solids and solutions. Uncertainty in quantitative XRF comes from sample preparation, background intensity (i.e. blanks), measurement of peak intensity, matrix effects, and the calibration curve. Typical reported limits of detection for XRF account only for background intensity (see 11), and are therefore not comparable with the uncertainty we describe here on ICP-MS measurements, which report the uncertainty as the difference between measured certified reference materials on the instrument to their certified values. This is a measure that can be termed global analytical uncertainty, and thus incorporates both random uncertainties that reduce the precision of an analytical result, and systematic uncertainties that reduce its accuracy. Relative global analytical uncertainty in XRF measurements on soil certified reference materials in literature are commonly much higher than the global analytical uncertainty achievable on ICP-MS, as demonstrated here: e.g. mean analytical uncertainty of 10.6%, 8.7%, 13.6%, and 19.2% for MgO, CaO, $Na_2O$, and $TiO_2$ respectively for 8 soil standards measured in triplicate on Philips MagiX Pro wavelength dispersive XRF (12); or 5.3% and 5.7% for Ca and Ti respectively for 9 soil standards on Innov-X Alpha 4000 XRF (13).

**1.8 Calculating ERW and CDR rates from leachate, soil exchangeable fraction and plant samples**

We employed elemental budgeting for all mesocosms in the experiment. Total amounts of Ca, Mg, and Na in all major non-solid phase pools were calculated: leachate (multiplying the concentration of cations in leachate by total volume of leachate, for each of up to six separate leachate events for each mesocosm), plant tissues (multiplying concentrations of cations in shoots, roots and seeds that underwent tissue digests by the respective mass of each tissue type collected from each mesocosm), and the soil exchangeable fraction (multiplying the concentration of cations in ammonium acetate leaches performed by the dry mass of soil at each separate depth interval for which leaches were performed).

To calculate a rate of weathering based on the additional flux of cations into non-soil pools for basalt-amended mesocosms, we subtracted from the total elemental budgets for Ca, Mg, and Na of each basalt-amended mesocosm the total elemental budget of the control mesocosm for each fertilizer scheme (NPK or manure) that in topsoil major element composition most closely matched the initial soil baseline. These were also the control mesocosms for each fertilizer scheme that yielded the smallest budget of cations in non-soil pools. Given variability in the size of elemental budgets between mesocosms of the same treatment, we report error on our calculations for cation loss as propagated standard deviations of all concentration measurements made for each treatment (NPK only, NPK + basalt, manure only, and manure + basalt). To convert from the additional flux of cations in moles to an amount of $CO_2$ consumed, we apply a modified Steinour formulation (as discussed in *Supporting Information Section 1.3*).

**1.9 Calculating ERW and CDR rates using the TiCAT theoretical scheme**

*Main Text Section 2.1* describes the conceptual framework of the mass-balance approach used to calculate weathering rates from solid-phase samples of the soil + basalt mixture, after basalt application.

Canonical approaches for estimating extent of weathering in natural systems (e.g., 14-26) define a mass transfer function for a mobile element $j$, $\tau^j$, such that:

$$\tau_w^j = \frac{m_{flux}^j}{m_p^j} = \left(\frac{c_w^j}{c_w^i} \middle/ \frac{c_p^j}{c_p^i}\right) - 1 \quad , \qquad (S3)$$

where $m$ is mass; $c$ is concentration; $p$ is unweathered parent material; $w$ is weathered material (and thus *flux* refers to mass lost from parent material during weathering); and $i$ is a reference immobile element (15).

Here, we use titanium as this reference immobile element. Titanium is commonly used as a reference element for immobility during weathering processes. This is due to its low solubility and the chemical stability of minerals within which it is abundant (e.g., rutile, titanite). Moreover, upon initial weathering of titanium-rich minerals, Ti tends to form insoluble secondary minerals. Thus, Ti is seldom removed from the solid phase (see 17, 27-31). A similar approach could be used with other largely immobile elements such as zirconium (Zr), thorium (Th), niobium (Nb) and tantalum (Ta) (see 17, 31-35). Aluminum (Al) and Rare Earth Elements are, in many cases, also immobile (e.g., 34, 36-40). The choice of immobile reference element should also depend on the composition of the soil onto which the feedstock is applied, and whether any fertilizers or other additives also contain this element.

Here, we focus on Ti due to its higher abundance in the mafic minerals that constitute the best ERW feedstocks, as well as the less onerous process of dissolving Ti for laboratory analysis compared to Zr (41). It is important to note, however, that the assumption that Ti acts as an immobile element during weathering must be treated with caution, as in rare cases Ti, Zr and Th — similar to Al and REEs — act as semi-mobile elements during weathering. However, for Ti to act as a semi-mobile element, extended periods of weathering in which the soil becomes extensively depleted are required. This contrasts strongly to the conditions in this experiment, or those that will prevail during ERW deployments on managed croplands (31, 35, 42-46).

We use the immobile behavior of titanium to compare the expected amount of basalt-derived cation, $CAT_{add}$, in a soil + basalt sample to the observed amount of cation in the sample, $CAT_{end}$, and calculate a fraction of basalt dissolution, $F_D$ (see *Main Text Section 2.1*). In the canonical mass transfer framework: $c_w^j = [CAT]_{end}$ ; $c_w^i = [Ti]_{end}$ ; $c_p^j = [CAT]_{add} + [CAT]_s$ ; $c_p^i = [Ti]_{add} + [Ti]_s$ ; thus:

$$F_D = -\tau_w^j \left( \frac{[CAT]_{add}}{[CAT]_s} + 1 \right) \quad . \quad (S4)$$

Note that here, $F_D$ of basalt specifically pertains to each cation of interest. For the purposes of estimating CDR the major cations in basalt are $Na^+$, $K^+$, $Ca^{2+}$, and $Mg^{2+}$. We disregard $K^+$ here, both because it is a constituent of the NPK fertilizer added to some of our mesocosm experiments, and because it represents only a minor portion of the overall cation load.

In this study, we use an extension of this scheme to calculate $F_D$ for each major cation: we define an additional endmember representing the fraction of basalt, $c_d$, that has been dissolved between the point of basalt application and the point of sampling. Thus, we resolve samples with composition $c_{end}$ as a mixture of three components; (i) basalt feedstock ($c_b$), (ii) soil baseline ($c_s$), and (iii) dissolved basalt + soil ($c_d$). $c_d$ comprises the Ti concentration of the basalt endmember, but retains only those mobile elements present in the soil baseline (i.e. for all CAT, $c_d = c_s$). This endmember therefore represents the loss of mobile base cations from the weathered basalt feedstock, and the buildup in the soil + basalt mixture of the immobile elements from this feedstock.

This scheme allows us to solve algebraically for all mixtures that satisfy the relationship $c_{end} = c_s + c_b + c_d$ (where $c_{end}$ is the sample composition). We can then calculate $F_D$ as the ratio of the mass fraction of dissolved basalt, $x_d$, to the mass fraction of total basalt in the sample at the point of application, $x_d + x_b$, such that:

$$F_D = \frac{x_d}{x_b + x_d} \quad . \quad (S5)$$

Plotting these three endmembers defines a region in [CAT] vs [Ti] space that can be thought of as a "mixing triangle", wherein all mixtures of soil + basalt, including those where weathering of basalt has occurred, should lie (**Fig. S6**).

The proportion of the three components that make up any soil + basalt sample within the "mixing triangle" can be solved algebraically, given a set of three linear equations that (a) describe the bulk composition of a sample as the mass-fraction-weighted average of each component (for Ti and any CAT) and (b) enforce mass conservation. These are given as:

$$C_{end}^i = x_s\, c_s^i + x_b\, c_b^i + x_d\, c_d^i \quad . \quad (S6)$$
$$C_{end}^j = x_s\, c_s^j + x_b\, c_b^j + x_d\, c_d^j \quad . \quad (S7)$$
$$1 = x_s + x_b + x_d \quad . \quad (S8)$$

where *i* represents immobile and *j* represents cation. The $x_{[\cdot]}$ and $c_{[\cdot]}$ terms represent the mass fraction and concentration of the soil, basalt and dissolution endmembers which are labelled by *s*, *b* and *d* subscripts, respectively. The $C_{end}$ terms represent the bulk composition of the detrital element or cation in the soil-basalt mixture after dissolution.
Then:

$$F_D = \frac{x_d}{x_b + x_d} \quad . \quad (S9)$$

Note that in order for these equations to have physical solutions (i.e. for each component to have a positive value), all $C_{end}$ values must plot within the area defined as the "mixing triangle". Several cases signify a breakdown of this scheme (**Fig. S6**):

1. Where a sample $C_{end}$ plots above the soil – basalt mixing line, such that **Δ[CAT] < 0** (**Fig. S6a**), this suggests that precipitation of a secondary mineral phase or phases containing cations has occurred between application of basalt and post-application sampling. This is pertinent to calculating initial CDR rates, as $CO_2$ is released by these reactions (with the exception of salt formation). In this study, we treat such samples as follows: (a) resolve $C_{end}^j$ for the base cation in question onto the soil-basalt mixing line using $C_{end}^i$ for that sample (such that $c_d = 0$); (b) calculate the difference in [CAT] between the original $C_{end}^j$ and the new $C_{end}^j$ value; and (c) calculate from this cation gain a resultant $CO_2$ release per unit area (see below).

2. Where a sample $C_{end}$ plots to the left of the "mixing triangle", such that **[Ti]end < [Ti]s** (**Fig. S6b**), this indicates that the soil baseline used is unsuitable, assuming no mass transport has occurred, and as such a more suitable soil baseline is required to obtain a reliable estimate of feedstock dissolution. This is not the case for any samples in this study.

3. Where a sample $C_{end}$ plots below the soil baseline, such that **[CAT]end < [CAT]s** (**Fig. S6c**), this indicates dissolution and loss of cations from the background soil. In this case it is necessary to apply the limit –Δ[CAT] = [CAT]add (i.e., $F_D$ = 1).

Given that the data for Ti and CAT are in the form of concentration (mass/mass), dissolution of basalt in a sample may result in a slight increase in concentration relative to the concentration of the same sample before dissolution has occurred, even if the amount of the element in question has not changed. This is because the total mass of the sample will be less after feedstock dissolution. Therefore, we apply the following correction to account for this effect:

$$F_D = \frac{(F_D^*) * (m_b * Ti_s - m_s * Ti_b)}{(F_D^* - 1) * m_b * Ti_s + m_b * Ti_s - m_s * Ti_b} \quad . \quad (S10)$$

where $F_D^*$ is $F_D$ calculated for each CAT before correcting for the concentration increase due to basalt dissolution, and $m_s$ and $m_b$ are the mass of soil and basalt respectively in an equivalent volume (taken here as the upper 12cm portion of each column). Note that amount Ti here is given as a mass, and not concentration. This formula is derived from first principles of mass balance. The correction factor $F_D/F_D^*$ is greater the smaller the difference between the concentration of the immobile tracer, in this case Ti, in the feedstock compared to the soil; and the smaller $F_D$. In our mesocosm experiment, this correction reduces $F_D$ by an average of 6% across all CAT and all replicates.

The TiCAT approach provides a time-integrated estimate of the amount of feedstock that has dissolved between application and post-application sampling. This can be directly converted into an initial CDR rate by using a modified Steinour formulation (*Supplemental Information Section 1.3*). Assuming $F_D^*$ for relevant cations CAT ($Na^+$, $Ca^{2+}$, $Mg^{2+}$) in each sample (see **Fig. S7**) is representative of $F_D^*$ for each mesocosm, we can calculate the total amount (mol) of each cation mobilized from applied feedstock over a given area, $n^j$ (mol m$^{-2}$):

$$n^j = F_D^* * A^j * \frac{10^{-3}}{M^j} \quad . \qquad (S11)$$

where $M^j$ is the molar mass of the cation, and $A^j$ is the application rate of the cation CAT (kg m$^{-1}$), calculated as:

$$A_b * \overline{c_b^j} = A^j \quad . \qquad (S12)$$

where $\overline{c_b^j}$ is the concentration of a cation in the basalt feedstock (mg kg$^{-1}$), and $A_b$ is the application rate of basalt, 5 kg m$^{-2}$ in this experiment.

To convert from an amount of cation dissolved, $n^j$, to an amount of $CO_2$ consumed during this weathering reaction, we apply the relevant Steinour coefficient (*see Supplemental Information Section 1.3*). Converting from this molar amount to a mass in kg m$^{-2}$ gives the total potential $CO_2$ removed by weathering of the feedstock for each cation; summed, these give total $CO_2$ removed by weathering of the feedstock in kg m$^{-2}$.

In our experimental method we remove the exchangeable fraction (non-mineral-bound cations occupying exchange sites on clay mineral surfaces). However, clay and carbonate minerals are included in the solid phase of digested soil samples, which could result in an underestimate of the extent of silicate mineral dissolution given that secondary mineral formation could reincorporate mobile cations into the solid phase following dissolution. As a result, our technique should be considered to track net cation mobility from the system, including the impact of secondary phase formation.

We use as our $c_{soil}$ endmember a sample of bulk soil ("initial soil") taken before the experiment was started (see **Figs. S8 and S9**). The total amount of metal cations added to each mesocosm via irrigation was negligible (see **Fig. S4** and **Table S2**). Additionally, the concentration of Ti, Mg, and Na in the solid phase of manure added to half of the mesocosms was below that of the soil endmember (see *Dataset S1*), so we disregarded contributions from the manure to the $c_{end}$ sample mixture. Even for Ca, which was enriched in manure relative to initial soil, this is a conservative approach when using the TiCAT method.

### 1.10 Correction for weathering due to strong acids

We apply a correction for weathering by nitric acid sourced from fertilizers in the topsoil of basalt-treated mesocosms before comparing the results from TiCAT to the weathering product tracing methods. Ammonium from urea and diammonium phosphate in chemical fertilizers, and from organic nitrogen compounds in manure (see *Supporting Information Section 1.2*) can be assumed to be converted to a strong acid through nitrification, when not utilized by the plant. This strong acid contributes to weathering but does not lead to the generation of carbonate alkalinity. Therefore, an amount of $CO_2$, equivalent (by the same charge balance framework, the Steinour formulation, presented in *Supporting Information Section 1.3*) to the amount of acidity added to the system by the fertilizer, is removed from each basalt-treated mesocosm's TiCAT initial CDR estimate before we report this estimate.

We first calculate the total amount (mol) of $NH_4^+$ added to each mesocosm in both fertilizer treatments. For air-dried manure, we assume $NH_4$-N to $NO_3$-N ratio of 21:1 (for stockpiled cow manure, from ref. 47). We assume a plant nitrogen use efficiency (NUE) of 55% (a median value in control experiments of sorghum bicolor by ref. 48), and therefore multiply the amount of $NH_4^+$ by (1-NUE) to give the amount of $NH_4^+$ nitrified in the soil. We then convert this amount to total acidity (mol $H^+$) by assuming that 1 mol $NH_4^+$ gives 2 mol $H^+$, given the stoichiometry of the nitrification reaction:

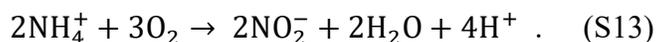

$$2NH_4^+ + 3O_2 \rightarrow 2NO_2^- + 2H_2O + 4H^+ \quad . \quad (S13)$$

The fertilizer-derived strong acidity added to each mesocosm was equivalent to $1.75 \times 10^4$ mol($H^+$) ha$^{-1}$ for the NPK treatment, and $1.53 \times 10^4$ mol($H^+$) ha$^{-1}$ for the manure treatment. Using a 1:1 molar ratio for $H^+$:$CO_2$eq converted to bicarbonate, we subtract the relevant amount of acidity from the TiCAT-derived estimate of initial CDR for each mesocosm. This is equivalent to 0.77 $tCO_2$eq ha$^{-1}$ for each NPK-treated mesocosm, and 0.67 $tCO_2$eq ha$^{-1}$ for each manure-treated mesocosm. When calculating ERW and CDR rates by using elemental budgets from leachate, soil exchangeable fraction and plant samples, a strong acid correction does not need to be used, as control mesocosms had the same quantities of fertilizers added as basalt-treated mesocosms.

Applying this correction, we see a significant reduction in the CDR we calculate from our mesocosm experiments: -28.2% on average for NPK-treated mesocosms, -44.3% on average for manure-treated mesocosms, and -34.4% on average across both fertilizer treatments. It is worth noting that in field settings a large portion of strong acids will react with bicarbonate and drive $CO_2$ evasion. We therefore are not implying that in all cases strong acid driven weathering should be factored into CDR estimates.

In our final initial CDR estimates (see Main Text Section 3, and Fig. 4), we report error as ± standard error of means, defined as the standard deviation divided by the square root of *n*, in this case 7 for each treatment type. Thus, this measure of uncertainty represents the spread in the weathering rates calculated for each mesocosm. Another way of presenting uncertainty here is to use propagated analytical uncertainty from the calculation of each weathering rate. Using this approach, our final initial CDR estimates are: means for both treatments of $1.47 \pm 0.15$ $tCO_2$eq ha$^{-1}$ (TiCAT), and $1.36 \pm 0.31$ $tCO_2$eq ha$^{-1}$ (dissolved pools); means for the NPK treatment of $1.96 \pm 0.22$ $tCO_2$eq ha$^{-1}$ (TiCAT), and $1.68 \pm 0.24$ $tCO_2$eq ha$^{-1}$ (dissolved pools); and means for the manure treatment of $0.97 \pm 0.20$ $tCO_2$eq ha$^{-1}$ (TiCAT), and $1.05 \pm 0.58$ $tCO_2$eq ha$^{-1}$ (dissolved pools).

## 2 Supporting Tables

**Table S1. Chemical composition (major oxides) of basalt feedstock used, Columbia River Basalt, Prineville Chemical Type Unit**. Concentration of Ca, Mg, Na, K and Ti from bulk rock total digest (HF+HNO$_3$+HCl), measured on Element HR-ICP-MS; other elements from XRF characterization. See also Kelland et al., 2020, and Lewis et al., 2021, for further chemical and textural information about the feedstock used.

| Oxide | Concentration (wt%) |
|---|---|
| CaO | 5.3 |
| MgO | 2.4 |
| Na$_2$O | 3.8 |
| K$_2$O | 3.0 |
| TiO$_2$ | 2.4 |
| Al$_2$O$_3$ | 13.0 |
| P$_2$O$_5$ | 2.3 |
| FeO | 9.7 |
| SiO$_2$ | 44.3 |

**Table S2. Chemical composition of artificial rainwater irrigation solution.**

| Salt | Solution concentration (mg L$^{-1}$) |
|---|---|
| NaNO$_3$ | 0.000135 |
| NaCl | 0.004790 |
| CaCl$_2$.2H$_2$O | 0.001100 |
| K$_2$SO$_4$ | 0.000450 |
| MgSO$_4$.7H$_2$O | 0.002050 |
| (NH$_4$)$_2$SO$_4$ | 0.002270 |
| pH | 5.56 |

## 3 Supporting Figures

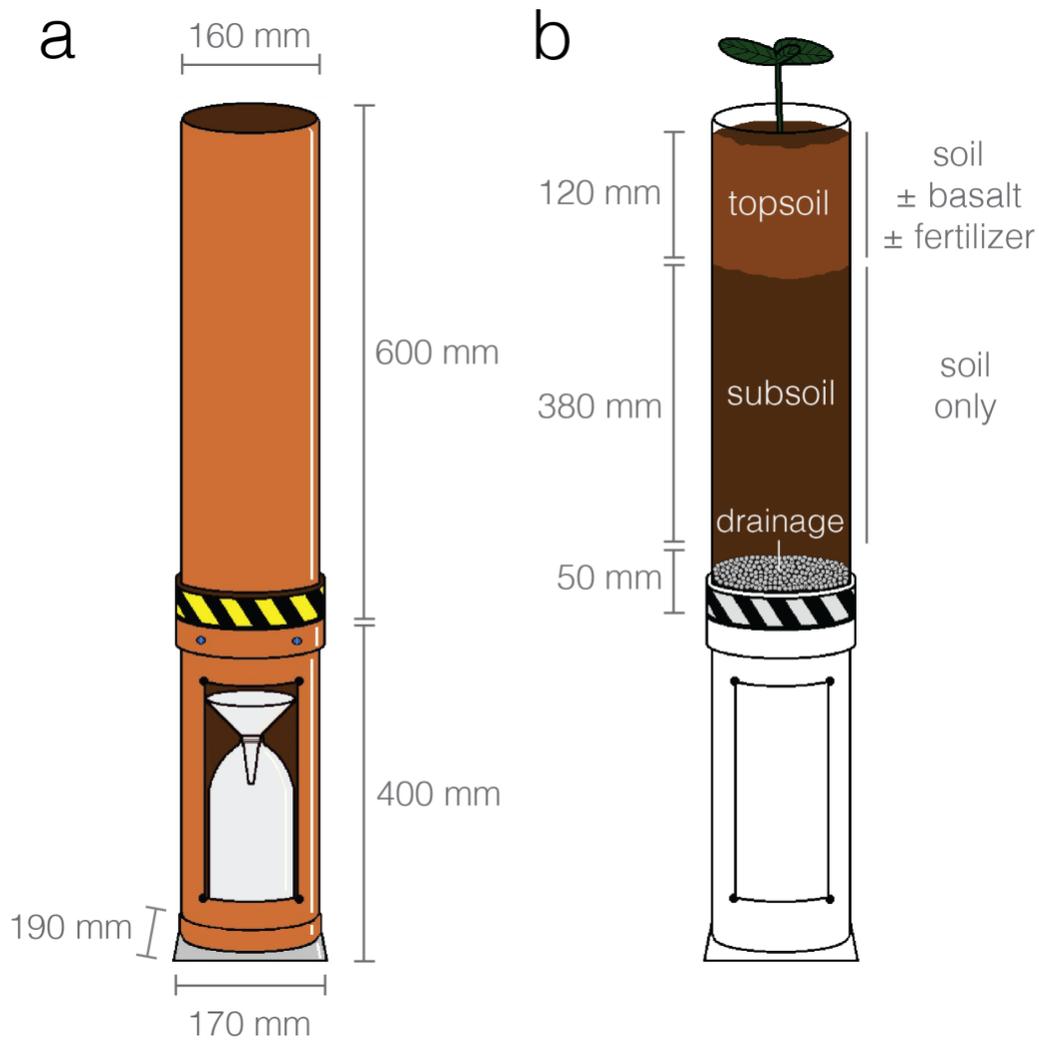

**Figure S1. Mesocosm design. (a)** Schematic of mesocosm exterior; and **(b)** cutaway indicating sub- and top-soil arrangement simulating tillage regime.

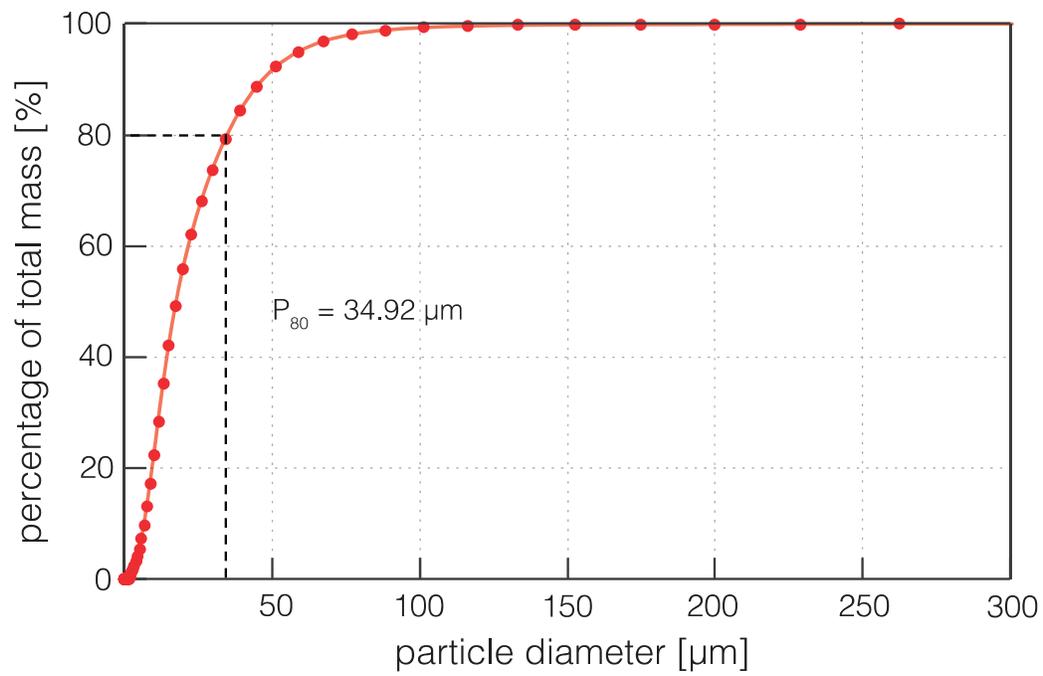

**Figure S2. Cumulative particle size distribution of basalt feedstock.** Determined by laser diffraction particle size distribution analysis.

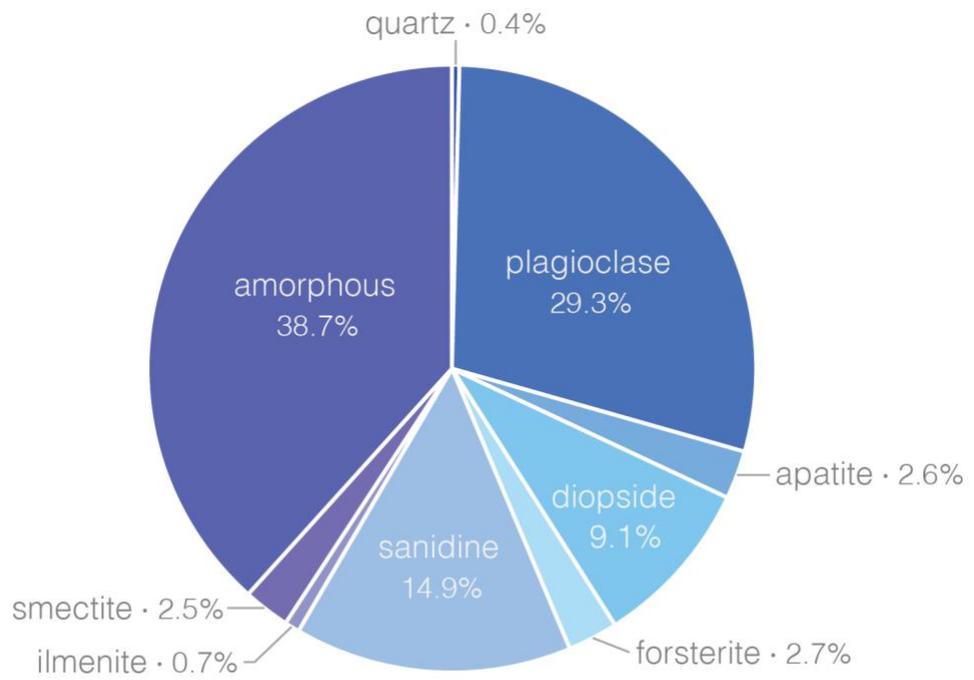

**Figure S3. Mineralogy of basalt feedstock.** Determined by X-Ray diffraction analysis.

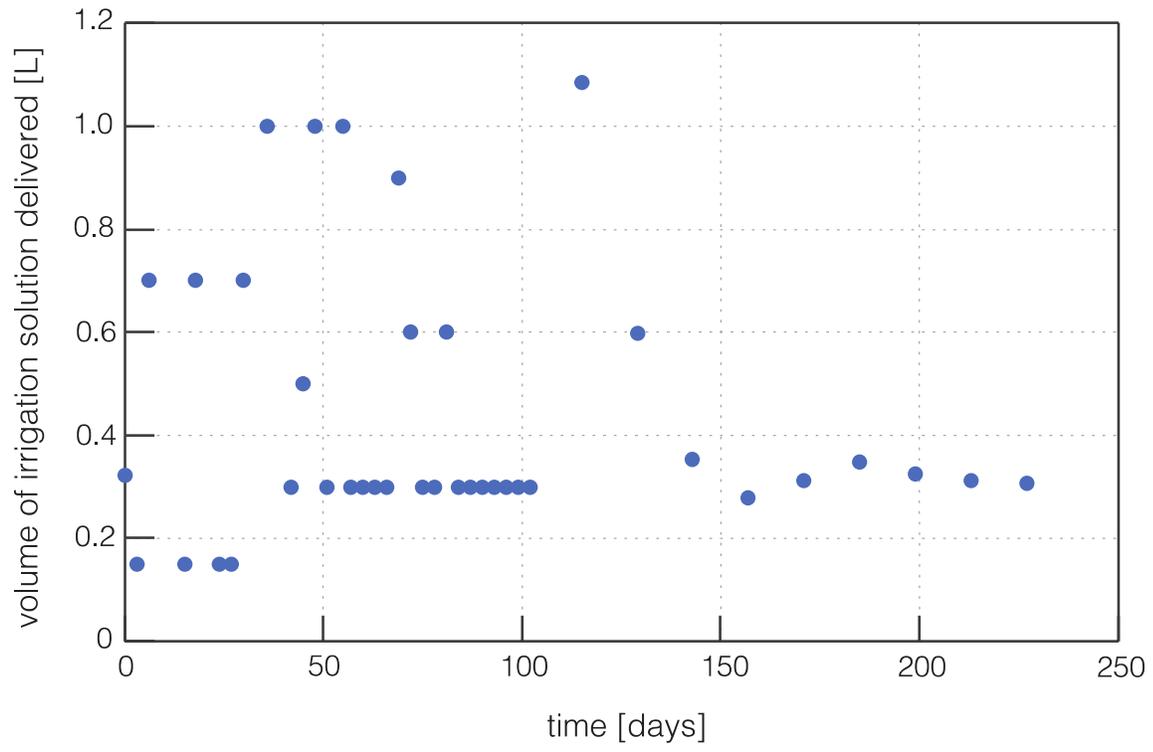

**Figure S4. Irrigation schedule.** Volume of irrigation solution added to mesocosms.

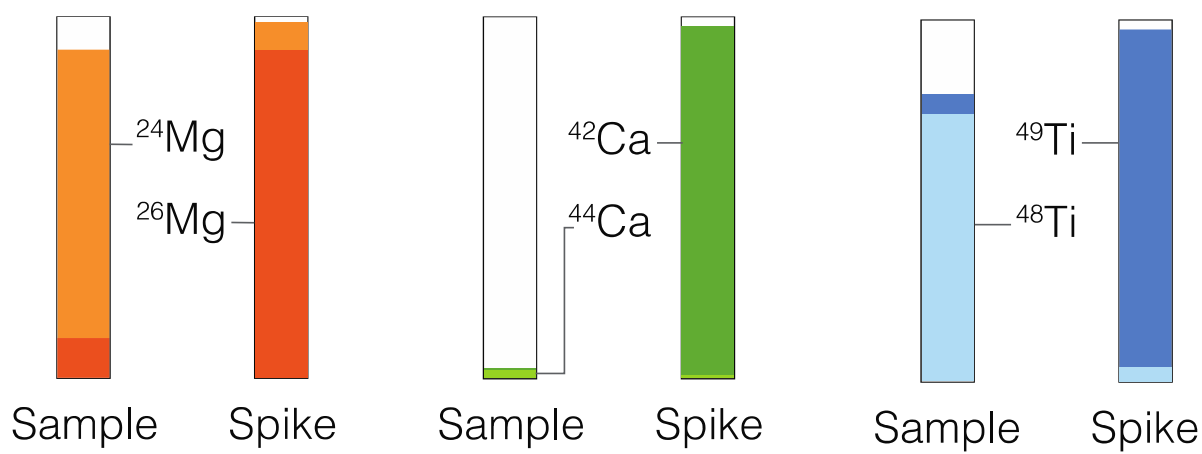

**Figure S5. Isotopic composition of sample and spike solutions for Mg, Ti, and Ca.** Measured isotopes of key elements for which isotope dilution was used to reduce analytical error in this study, in natural samples (left) and the isotope spike cocktail added to the samples (right). Isotopes shown are those used in calculations. Note that $^{40}$Ca is the most abundant stable isotope of Ca in natural samples, but is not measured due to interference on ICP-MS from $^{40}$Ar.

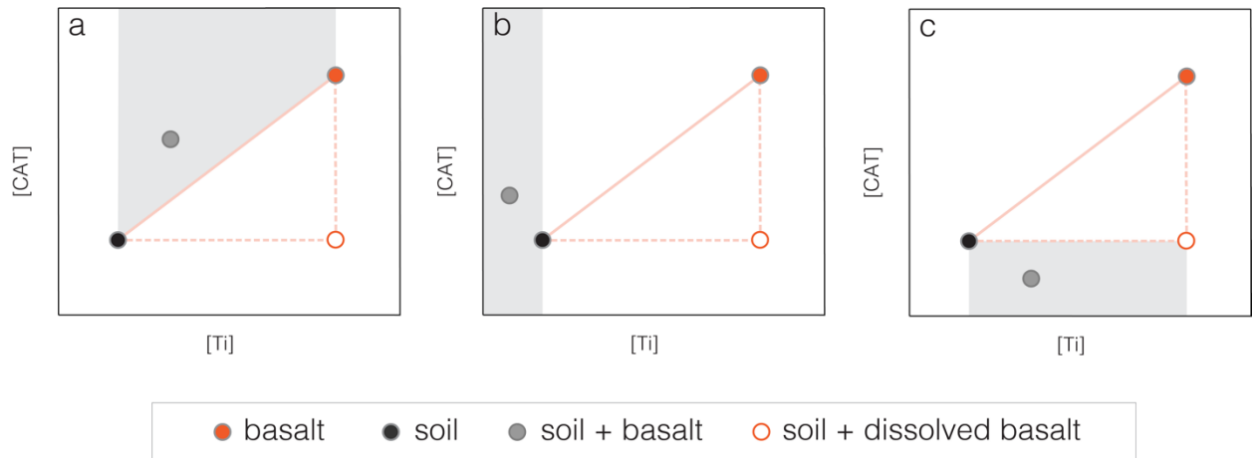

**Figure S6. The TiCAT conceptual framework as a three-component "mixing triangle".** The endmember soil + dissolved basalt ($c_d$) has the Ti composition of the basalt endmember and the CAT composition of the soil endmember. (a-c) show cases where the composition of soil + basalt mixtures fall outside of the "mixing triangle" and therefore cannot be resolved in this framework: $\Delta$CAT < 0, indicating precipitation of CAT into a secondary mineral phase (a); $Ti_{end} < Ti_{soil}$, indicating the soil baseline value for Ti is not suitable (b); $CAT_{end} < CAT_{soil}$, indicating dissolution of CAT from soil (c).

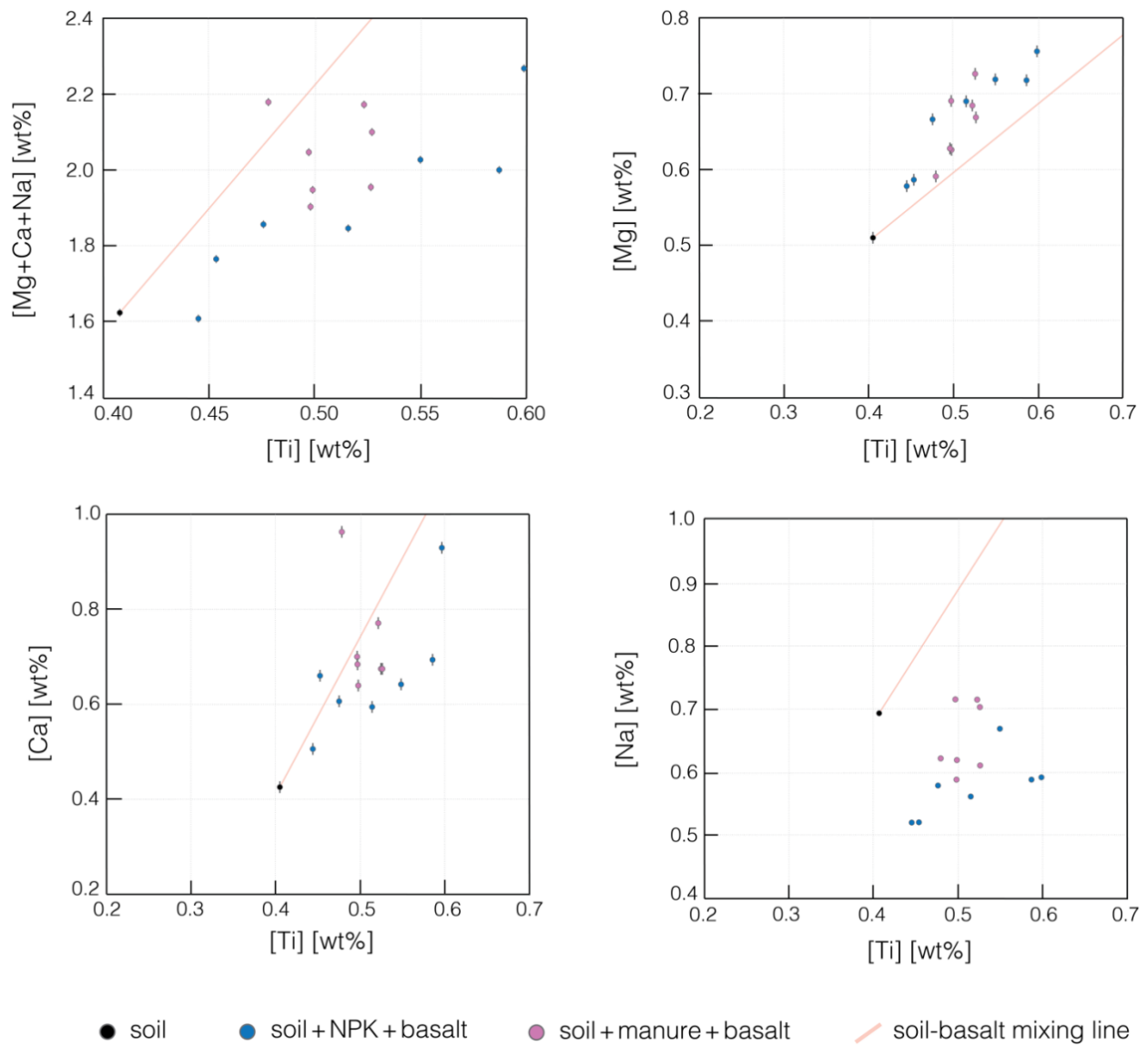

**Figure S7. Concentration in ashed solid samples of Mg, Ca, and Na from the "topsoil" depth interval (0-12 cm) in all basalt-amended mesocosms in this study** (n=14). "Soil" is a sample of initial soil that mesocosms were filled with. Analytical error is shown (note the error bar for Na is smaller than the symbol).

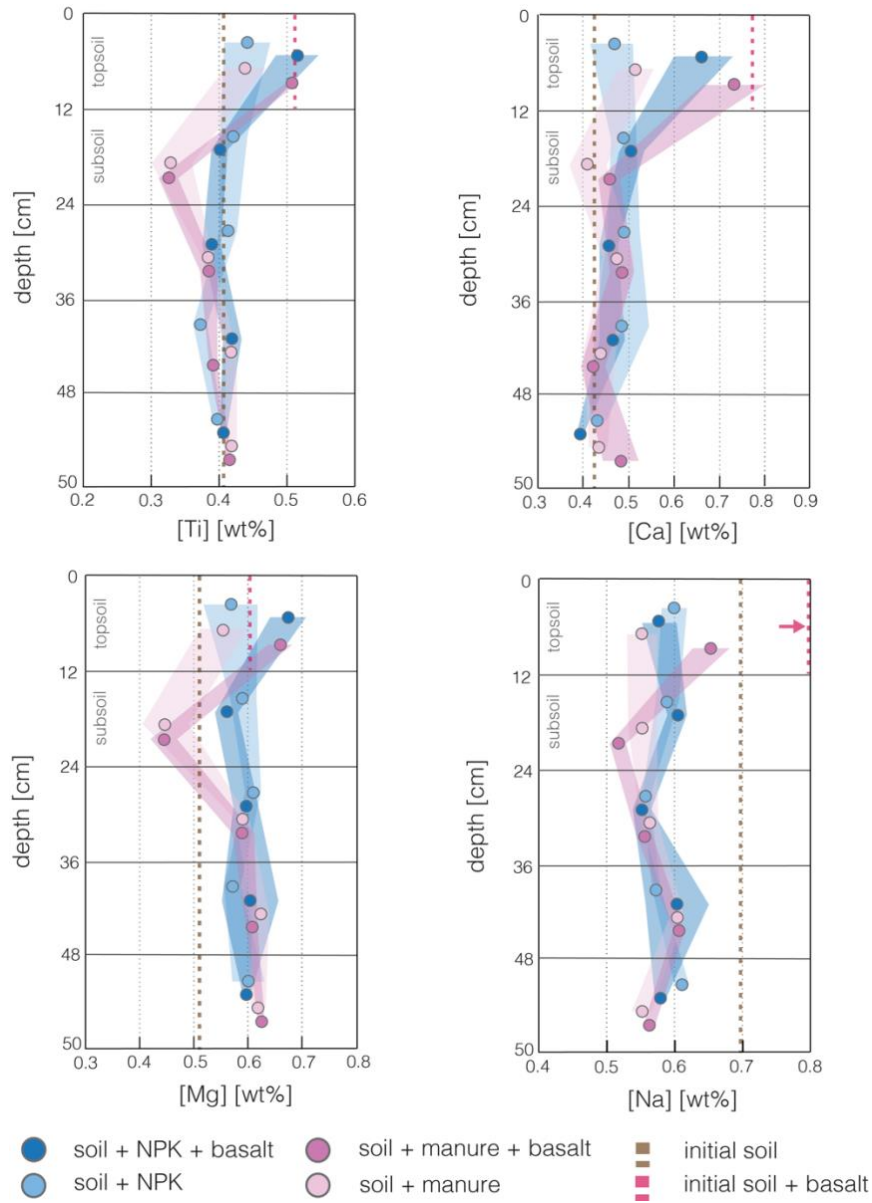

**Figure S8. Depth profiles of concentration in ashed solid samples of Ti, Ca, Mg, and Na from mesocosms in this study**. Points are averages of all columns (n=7) for the relevant treatment, and shaded envelopes show σ. Samples are all homogenized across the depth interval between solid horizontal bars. Dashed brown vertical line indicates the composition of the initial soil endmember. Dashed red vertical line indicates the composition of a fully homogenized initial soil + basalt mixture in the topsoil layer, where the average additional amount of Ti is used to calculate average basalt addition rate in samples.

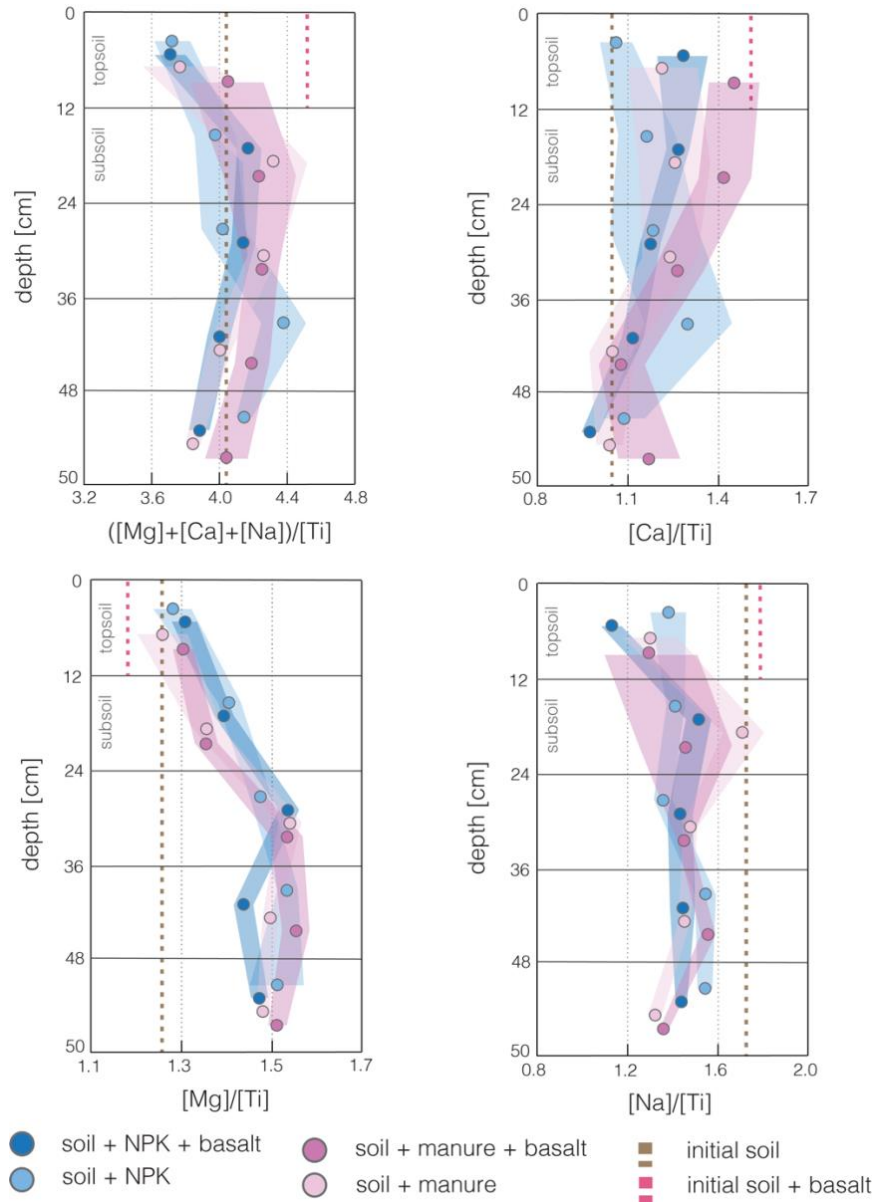

**Figure S9. Depth profiles of ratios in concentration in ashed solid samples of (Ca+Mg+Na)/Ti, Ca/Ti, Mg/Ti, and Na/Ti from mesocosms in this study.** Points are averages of all columns (n=7) for the relevant treatment, and shaded envelopes show 2σ. Samples are all homogenized across the depth interval between solid horizontal bars. Dashed brown vertical line indicates the composition of the initial soil endmember. Dashed red vertical line indicates the composition of a fully homogenized initial soil + basalt mixture in the topsoil layer, where the average additional amount of Ti is used to calculate average basalt addition rate in samples.

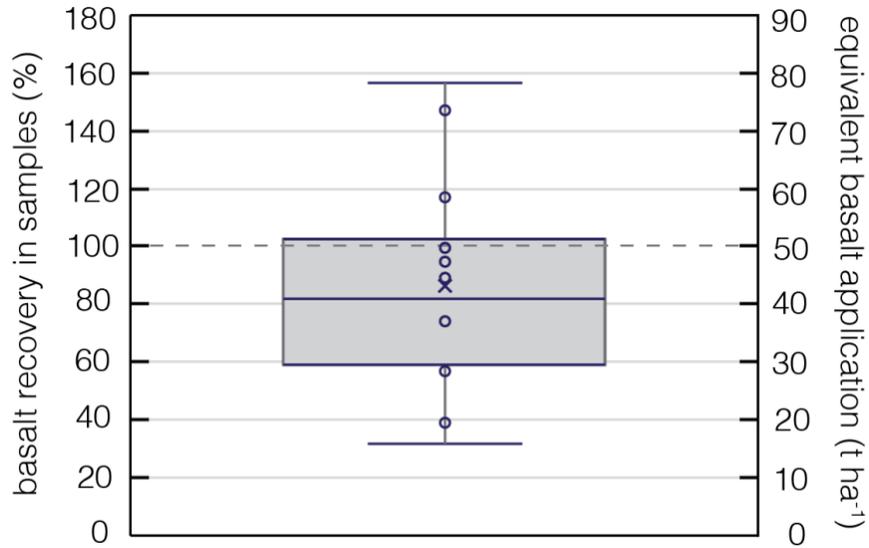

**Figure S10. Box and whisker plot showing % recovery of basalt feedstock in samples from all basalt-treated mesocosms** (n=14), as determined by the amount of titanium in samples. 100% basalt recovery means the amount of titanium in a sample is equal to the expected amount of titanium, proportional to the total addition of basalt to the topsoil and homogeneous mixing to 12cm. The right-hand axis shows the equivalent basalt application that, when homogeneously mixed into the topsoil layer, results in the amount of titanium present in samples. X shows the mean.

# 4 Supporting References

1. Kelemen, P. B.; McQueen, N.; Wilcox, J.; Renforth, P.; Dipple, G.; Vankeuren, A. P. Engineered Carbon Mineralization in Ultramafic Rocks for $CO_2$ Removal from Air: Review and New Insights. *Chem Geol* **2020**, *550*, 119628. https://doi.org/10.1016/j.chemgeo.2020.119628.
2. Lewis, A. L.; Sarkar, B.; Wade, P.; Kemp, S. J.; Hodson, M. E.; Taylor, L. L.; Yeong, K. L.; Davies, K.; Nelson, P. N.; Bird, M. I.; Kantola, I. B.; Masters, M. D.; DeLucia, E.; Leake, J. R.; Banwart, S. A.; Beerling, D. J. Effects of Mineralogy, Chemistry and Physical Properties of Basalts on Carbon Capture Potential and Plant-Nutrient Element Release via Enhanced Weathering. *Appl Geochem* **2021**, *132*, 105023. https://doi.org/10.1016/j.apgeochem.2021.105023.
3. H. H. Steinour, Some effects of carbon dioxide on mortars and concrete-discussion. *J. Am. Concrete*, **1959**, *I 30*, 905
4. M. Fernández Bertos; S. J. R. Simons; C. D. Hills, P. J. Carey. A review of accelerated carbonation technology in the treatment of cement-based materials and sequestration of $CO_2$. *J. Hazard Mater,* **2004**, *112*, 193–205.
5. Gunning, P. J.; Hills, C. D.; Carey, P. J. Accelerated Carbonation Treatment of Industrial Wastes. *Waste Manage* **2010**, *30* (6), 1081–1090. https://doi.org/10.1016/j.wasman.2010.01.005.
6. Renforth, P. The Negative Emission Potential of Alkaline Materials. *Nat Commun* **2019**, *10* (1), 1401. https://doi.org/10.1038/s41467-019-09475-5.
7. Bullock, L. A.; James, R. H.; Matter, J.; Renforth, P.; Teagle, D. A. H. Global Carbon Dioxide Removal Potential of Waste Materials From Metal and Diamond Mining. *Frontiers Clim* **2021**, *3*, 694175. https://doi.org/10.3389/fclim.2021.694175.
8. Rinder, T.; von Hagke, C. The Influence of Particle Size on the Potential of Enhanced Basalt Weathering for Carbon Dioxide Removal - Insights from a Regional Assessment. *J Clean Prod* **2021**, *315*, 128178. https://doi.org/10.1016/j.jclepro.2021.128178.
9. Renforth, P. The Potential of Enhanced Weathering in the UK. *Int J Greenh Gas Con* **2012**, *10*, 229–243. https://doi.org/10.1016/j.ijggc.2012.06.011.
10. Chapman, H.D. Cation-exchange capacity. In: *Methods of soil analysis - Chemical and microbiological properties*. C.A. Black; ed. Agronomy **1965** .9:891-901.
11. Rousseau, R. M. Detection Limit and Estimate of Uncertainty of Analytical XRF Results. *Rigaku J*. **2001**, *18* (2), 33-47.
12. Krishna, A. K.; Murthy, N. N.; Govil, P. K. Multielement Analysis of Soils by Wavelength-Dispersive X-Ray Fluorescence Spectrometry. *Atom. Spectrosc*. **2007**, *28* (6), 202-214.
13. Kenna, T. C.; Nitsche, F. O.; Herron, M. M.; Mailloux, B. J.; Peteet, D.; Sritrairat, S.; Sands, E.; Baumgarten, J. Evaluation and Calibration of a Field Portable X-Ray Fluorescence Spectrometer for Quantitative Analysis of Siliciclastic Soils and Sediments. *J. Anal. At. Spectrom.* **2010**, *26* (2), 395–405. https://doi.org/10.1039/c0ja00133c.
14. Brimhall, G.H.; Dietrich, W.E.. Constitutive mass balance relations between chemical composition, volume, density, porosity, and strain in metasomatic hydrochemical systems: Results on weathering and pedogenesis. *Geochim Cosmochim Ac*, **1987**, *51*, 567–587. https://doi.org/10.1016/0016-7037(87)90070-6

15. Chadwick, O.A.; Brimhall, G.H.; Hendricks, D.M.. From a black to a gray box — a mass balance interpretation of pedogenesis. *Geomorphology*, **1990**, *3*, 369–390. https://doi.org/10.1016/0169-555x(90)90012-f
16. Chadwick, O.A.; Derry, L.A.; Vitousek, P.M.; Huebert, B.J.; Hedin, L.O.. Changing sources of nutrients during four million years of ecosystem development. *Nature*, **1999**, *397*, 491–497. https://doi.org/10.1038/17276
17. Brimhall, G.H.; J., L.C., Ford; C., Bratt, J.; Taylor, G.; Warin, O.. Quantitative geochemical approach to pedogenesis: importance of parent material reduction, volumetric expansion, and eolian influx in lateritization. *Geoderma*, **1991**, *51*, 51–91. https://doi.org/10.1016/0016-7061(91)90066-3
18. Kurtz, A.C.; Derry, L.A.; Chadwick, O.A.; Alfano, M.J.. Refractory element mobility in volcanic soils. *Geology*, **2000**, *28*, 683–686. https://doi.org/10.1130/0091-7613(2000)28<683:remivs>2.0.co;2
19. White, A.F.; Bullen, T.D.; Schulz, M.S.; Blum, A.E.; Huntington, T.G.; Peters, N.E.. Differential rates of feldspar weathering in granitic regoliths. *Geochim Cosmochim Ac,* **2001**, *65*, 847–869. https://doi.org/10.1016/s0016-7037(00)00577-9
20. Anderson, S.P., Dietrich, W.E., Brimhall, G.H., 2002. Weathering profiles, mass-balance analysis, and rates of solute loss: Linkages between weathering and erosion in a small, steep catchment. Gsa Bulletin 114, 1143–1158. https://doi.org/10.1130/0016-7606(2002)114<1143:wpmbaa>2.0.co;2
21. Riebe, C.S., Kirchner, J.W., Finkel, R.C., 2003. Long-term rates of chemical weathering and physical erosion from cosmogenic nuclides and geochemical mass balance. Geochim Cosmochim Ac 67, 4411–4427. https://doi.org/10.1016/s0016-7037(03)00382-x
22. Tabor, N.J., Montañez, I.P., Zierenberg, R., Currie, B.S., 2004. Mineralogical and geochemical evolution of a basalt-hosted fossil soil (Late Triassic, Ischigualasto Formation, northwest Argentina): Potential for paleoenvironmental reconstruction. Gsa Bulletin 116, 1280–1293. https://doi.org/10.1130/b25222.1
23. Sheldon, N.D., Tabor, N.J., 2009. Quantitative paleoenvironmental and paleoclimatic reconstruction using paleosols. Earth-sci Rev 95, 1–52. https://doi.org/10.1016/j.earscirev.2009.03.004
24. Brantley, S.L., Lebedeva, M., 2011. Learning to Read the Chemistry of Regolith to Understand the Critical Zone. Annu Rev Earth Pl Sc 39, 387–416. https://doi.org/10.1146/annurev-earth-040809-152321
25. Fisher, B.A., Rendahl, A.K., Aufdenkampe, A.K., Yoo, K., 2017. Quantifying weathering on variable rocks, an extension of geochemical mass balance: Critical zone and landscape evolution. Earth Surf. Process. Landforms 42, 2457–2468. https://doi.org/10.1002/esp.4212
26. Lipp, A.G., Shorttle, O., Sperling, E.A., Brocks, J.J., Cole, D.B., Crockford, P.W., Mouro, L.D., Dewing, K., Dornbos, S.Q., Emmings, J.F., Farrell, U.C., Jarrett, A., Johnson, B.W., Kabanov, P., Keller, C.B., Kunzmann, M., Miller, A.J., Mills, N.T., O'Connell, B., Peters, S.E., Planavsky, N.J., Ritzer, S.R., Schoepfer, S.D., Wilby, P.R., Yang, J., 2021. The composition and weathering of the continents over geologic time. Geochem Perspectives Lett 21–26. https://doi.org/10.7185/geochemlet.2109

27. Monro, S.K., Loughnan, F.C., and Walker, M.C., 1983, The Ayrshire bauxitic clay: An allochthonous deposit?, in Wilson, R.C.L., ed., Residual deposits: Geological Society (London) Special Publication 11, p. 47–58
28. Middelburg, J.J., Weijden, C.H. van der, Woittiez, J.R.W., 1988. Chemical processes affecting the mobility of major, minor and trace elements during weathering of granitic rocks. Chem Geol 68, 253–273. https://doi.org/10.1016/0009-2541(88)90025-3
29. Tilley, D.B., Eggleton, R.A., 2005. Titanite Low-Temperature Alteration and Ti Mobility. Clay Clay Miner 53, 100–107. https://doi.org/10.1346/ccmn.2005.0530110
30. Taboada, T., Cortizas, A.M., García, C., García-Rodeja, E., 2006. Particle-size fractionation of titanium and zirconium during weathering and pedogenesis of granitic rocks in NW Spain. Geoderma 131, 218–236. https://doi.org/10.1016/j.geoderma.2005.03.025
31. Du, X., Rate, A.W., Gee, M.A.M., 2012. Redistribution and mobilization of titanium, zirconium and thorium in an intensely weathered lateritic profile in Western Australia. Chem Geol 330, 101–115. https://doi.org/10.1016/j.chemgeo.2012.08.030
32. Langmuir, D., Herman, J.S., 1980. The mobility of thorium in natural waters at low temperatures. Geochim Cosmochim Ac 44, 1753–1766. https://doi.org/10.1016/0016-7037(80)90226-4
33. Braun, J.-J., Pagel, M., Herbilln, A., Rosin, C., 1993. Mobilization and redistribution of REEs and thorium in a syenitic lateritic profile: A mass balance study. Geochim Cosmochim Ac 57, 4419–4434. https://doi.org/10.1016/0016-7037(93)90492-f
34. Little, M.G., Lee, C.-T.A., 2010. Sequential extraction of labile elements and chemical characterization of a basaltic soil from Mt. Meru, Tanzania. J Afr Earth Sci 57, 444–454. https://doi.org/10.1016/j.jafrearsci.2009.12.001
35. Jiang, K., Qi, H.-W., Hu, R.-Z., 2018. Element mobilization and redistribution under extreme tropical weathering of basalts from the Hainan Island, South China. J Asian Earth Sci 158, 80–102. https://doi.org/10.1016/j.jseaes.2018.02.008
36. Nesbitt, H.W., 1979. Mobility and fractionation of rare earth elements during weathering of a granodiorite. Nature 279, 206–210. https://doi.org/10.1038/279206a0
37. Gouveia, M.A., Prudêncio, M.I., Figueiredo, M.O., Pereira, L.C.J., Waerenborgh, J.C., Morgado, I., Pena, T., Lopes, A., 1993. Behavior of REE and other trace and major elements during weathering of granitic rocks, Évora, Portugal. Chem Geol 107, 293–296. https://doi.org/10.1016/0009-2541(93)90194-n
38. Boulangé, B., Colin, F., 1994. Rare earth element mobility during conversion of nepheline syenite into lateritic bauxite at Passa Quatro, Minais Gerais, Brazil. Appl Geochem 9, 701–711. https://doi.org/10.1016/0883-2927(94)90029-9
39. Öhlander, B., Land, M., Ingri, J., Widerlund, A., 1996. Mobility of rare earth elements during weathering of till in northern Sweden. Appl Geochem 11, 93–99. https://doi.org/10.1016/0883-2927(95)00044-5
40. Tertre, E., Hofmann, A., Berger, G., 2008. Rare earth element sorption by basaltic rock: Experimental data and modeling results using the "Generalised Composite approach." Geochim Cosmochim Ac 72, 1043–1056. https://doi.org/10.1016/j.gca.2007.12.015
41. Jain, J.C., Neal, C.R., Hanchar, J.M., 2001. Problems Associated with the Determination of Rare Earth Elements of a "Gem" Quality Zircon by Inductively Coupled Plasma-Mass Spectrometry. Geostandard Newslett 25, 229–237. https://doi.org/10.1111/j.1751-908x.2001.tb00598.x

42. Melfi, A.J., Subies, F., Nahon, D., Formoso, M.L.L., 1996. Zirconium mobility in bauxites of Southern Brazil. J S Am Earth Sci 9, 161–170. https://doi.org/10.1016/0895-9811(96)00003-x
43. Cornu, S., Lucas, Y., Lebon, E., Ambrosi, J.P., Luizão, F., Rouiller, J., Bonnay, M., Neal, C., 1999. Evidence of titanium mobility in soil profiles, Manaus, central Amazonia. Geoderma 91, 281–295. https://doi.org/10.1016/s0016-7061(99)00007-5
44. Hill, I.G., Worden, R.H., Meighan, I.G., 2000. Yttrium: The immobility-mobility transition during basaltic weathering. Geology 28, 923–926. https://doi.org/10.1130/0091-7613(2000)28<923:ytitdb>2.0.co;2
45. Hodson, M.E., 2002. Experimental evidence for mobility of Zr and other trace elements in soils. Geochim Cosmochim Ac 66, 819–828. https://doi.org/10.1016/s0016-7037(01)00803-1
46. Bednar, A.J., Gent, D.B., Gilmore, J.R., Sturgis, T.C., Larson, S.L., 2004. Mechanisms of Thorium Migration in a Semiarid Soil. J Environ Qual 33, 2070–2077. https://doi.org/10.2134/jeq2004.2070
47. Larney, F. J.; Buckley, K. E.; Hao, X.; McCaughey, W. P. Fresh, Stockpiled, and Composted Beef Cattle Feedlot Manure. *J Environ Qual* **2006**, *35* (5), 1844–1854. https://doi.org/10.2134/jeq2005.0440.
48. Wang, W.-F.; Zong, Y.-Z.; Zhang, S-Q. Water- and Nitrogen-use Efficiencies of Sweet Sorghum Seedlings are Improved under Water Stress. *Int. J. Agric. Biol.* **2014**, *16*, 285-292.